\documentclass[12pt]{article}
\usepackage{epsfig}

\begin{document}
\begin{center}
{\bf EXPECTATION VALUES OF FOUR-QUARK OPERATORS IN THE NUCLEON
}\\

\vspace{1cm}

E.G.~Drukarev, M.G.~Ryskin, V.A.~Sadovnikova\\
Petersburg Nuclear Physics Institute, St. Petersburg 188300, Russia;\\
V.E.~Lyubovitskij, Th.~Gutsche and Amand~Faessler\\
Institut f\"ur Theoretische Physik, Universit\"at T\"ubingen,\\
 T\"ubingen D-72076, Germany \\
\end{center}

\begin{abstract}
We calculate expectation values of  QCD
operators consisting of the
products of the four operators
of the light quarks $\bar q\Gamma^X q\bar
q\Gamma^Y q$, with $\Gamma^{X,Y}$ corresponding to the scalar,
pseudoscalar, vector, pseudovector (axial) and tensor Lorentz
structures, in the nucleon.  All  combinations of the light flavors are
considered.
For the evaluation we use elements of
 the Perturbative Chiral Quark Model (PCQM),
 approximating the contribution of the valence quarks by the
contribution of the PCQM constituent quarks.
The contribution of the sea quarks is treated by averaging
 over the QCD pions with the distribution of the
 pion  field being determined by the PCQM.
For quarks with the same flavor the expectation values
are dominated by the contribution of the sea-quarks.
In the scalar case the contribution of the sea quarks
is dominated by the "disconnected terms" were one of
the pairs of the $\bar qq$ operators acts on the vacuum while
the other
one acts on the quarks of the pion. The role of
the interference terms with  one of the $\bar qq$ pairs acting on
the sea quarks and another one acting on the valence quarks
increases for quarks with different flavors. The result
for the scalar condensate is compared to the one obtained
earlier in the framework of the Nambu -Jona-Lasinio
model.
\end{abstract}

\vspace{1.5cm}

\section{Introduction}

It is known that expectation values of the two-quark QCD
operators $\bar qq$ give the total number of quarks and antiquarks
in a hadron under certain reasonable assumptions \cite{a1,a2}.
The motivation for studies of the expectation values of the
four-quark operators in hadrons is that they carry information
about the correlation of $\bar qq$ pairs. These expectation values in
nucleons determine the coefficients of the next-to-leading order
operators of deep inelastic scattering \cite{s}.  One more
reason is the manifestation of such operators in QCD sum
rules in nuclear matter  \cite{96,119}. The lack of data on
the expectation values of these operators became one of the main
obstacles for further development of this approach.

Unfortunately, the calculations of these expectation values require
 some model assumptions on the quark structure of the nucleon. As it
stands now, the only calculation of the four-quark condensates in
the nucleon is the one carried out by Celenza et al.  \cite{44} for the
scalar case in the framework of the Nambu and Jona-Lasinio model  \cite{1}
under certain additional assumptions.  Also, some constraints on
the values were obtained by Johnson and Kisslinger  \cite{k1} by
analyzing QCD sum rules for nucleons and isobars.

In this paper we calculate the expectation values of the four-quark
 condensates in nucleons by using elements of
 the perturbative chiral
quark model (PCQM).  The chiral quark model, originally
suggested in
\cite{26}, was fully set up in
 \cite{l1, 25, 25-ad}. In the PCQM the nucleon is
 treated as a system of relativistic
valence quarks moving in an effective static field.
In addition, the valence quarks are supplemented by a perturbative cloud of
pseudoscalar mesons as dictated by chiral symmetry requirements.
In the current paper we restrict
to the
  simplest SU(2) version of the PCQM, which includes
only pions. To facilitate the evaluation of the four-quark operators we
resort to previously derived results \cite{25}, such as nucleon wave function
renormalization and self-energy contributions, which are used
as an input in the present derivation. Although we do not apply
the full perturbative machinery of the PCQM, as  laid out in \cite{l1}, the present
evaluation serves as a first indication for the values of the four-quark
condensates.

We calculate the
expectation values of the four-quark operators as the matrix elements
 of the PCQM
nucleon.
In general the QCD operators $q$ act on the valence and the sea quarks. We
approximate the averaging
 over the valence quarks by averaging over the PCQM
constituent quarks and also approximate the averaging over the sea quarks by
averaging over the pions.
In the original formulation
of the PCQM the pions were treated as separate degrees
of freedom, without taking into  account
 their quark structure. Actually, in \cite{25}
 the sigma-term was calculated
in the framework of PCQM, with the pions determining
the sea-quark contribution.
We shall employ an additional extension of the PCQM, considering
the pions as physical particles
 with their quark contents being determined
by QCD.
In this approach we therefore obtain
 the excess of the four-quark operator expectation
value over the one of the vacuum  in the nucleon volume.

In the following we calculate the expectation values
\begin{equation} \label{1n}
U^{XY,f_1f_2} = \langle N|T^{XY,f_1f_2}|N\rangle
\end{equation}
of the four-quark operators of  light quarks
\begin{equation} \label{2n}
T^{XY,f_1f_2} = (:\bar q^{f_1a}\Gamma^X q^{f_1a'} \cdot
\bar q^{f_2b}\Gamma^Y q^{f_2b'}:)
(\delta_{aa'}\delta_{bb'}-\delta_{ab'}\delta_{ba'}).
\end{equation}
Here $\Gamma^X$, $\Gamma^Y$  are the matrices acting on the Lorentz
indices, $q^f$ denote QCD quark operators with $f$ standing for the
flavor. The dots denote the normal ordering of the operators, $a,
a', b, b'$ represent the color indices. We calculate the
condensates for the basic $4\times 4$ matrices
\begin{equation} \label{gg}
\Gamma^S =I, \quad \Gamma^{Ps} = \gamma_5, \quad
\Gamma^{V}_\mu = \gamma_\mu, \quad \Gamma^A_\mu =
\gamma_\mu\gamma_5, \quad \Gamma^T_{\mu\nu} = \frac
i2(\gamma_\mu\gamma_\nu - \gamma_\nu\gamma_\mu)
\end{equation}
with $\gamma_\mu$ being the Dirac matrices, that is we
consider  the scalar, pseudoscalar,
vector, pseudovector (axial) and tensor cases. We also obtain the
results for the mixed condensate $U^{SV,du}$ which is important in
applications.

These are three types of contributions to $U^{XY,f_1f_2}$ in our
 approach.  All  four operators $q$ can act on the constituent quarks,
providing the term $C^{XY,f_1f_2}$. Also, four operators can act on
the pions providing the term $P^{XY,f_1f_2}$. There is also a
possibility that two of the operators act on the constituent quarks
while the other two act on the pions.  Denoting the last term as
$J^{XY,f_1f_2}$, we present the expectation values as
\begin{equation} \label{2}
U^{XY,f_1f_2}
= C^{XY,f_1f_2} + P^{XY,f_1f_2} + J^{XY,f_1f_2}.
\end{equation}
In Appendix A we show how these contributions manifest
themselves in the PCQM formalism.

To simplify the notations we introduce
\begin{equation} \label{2m}
U^{X,f_1f_2}= U^{XX,f_1f_2}
\end{equation}
with a similar convention for the other functions ($C, P, J, T$)
involved.

While the four-quark
condensates are Lorentz scalars in the scalar and
pseudoscalar channels, they have a more complicated structure in the
case of the vector and axial channels:
\begin{equation} \label{1m}
U^{V(A)}_{\mu\nu} = a^{V(A)}g_{\mu\nu} + b^{V(A)}\frac{p_\mu
p_\nu} {m^2}.
\end{equation}
Here $p_\mu$ is the momentum of the nucleon, while $m$ denotes the
nucleon mass. Also, in the tensor channel we have
\begin{equation} \label{1l}
U^T_{\mu\nu,\alpha\beta} = a^T s_{\mu\nu,\alpha\beta}
 + b^T t_{\mu\nu,\alpha\beta}
\end{equation}
with
\begin{equation} \label{9}
s_{\mu\nu,\alpha\beta} = g_{\mu\alpha} g_{\nu\beta} - g_{\mu\beta}
g_{\nu\alpha},
\end{equation}
$$
t_{\mu\nu,\alpha\beta} = \frac{1}{m^2}
(p_\mu p_\alpha g_{\nu\beta} + p_\nu p_\beta
g_{\mu\alpha} - p_\mu p_\beta g_{\nu\alpha} - p_\nu p_\alpha
g_{\mu\beta}).
$$
We shall denote the values of $a^{V,A,T}$ and $b^{V,A,T}$
corresponding to the contributions $C$, $P$ and $J$ by the lower
indices, i.e. $a^V_{C,P,J}$, etc.

We approximate the averaging
 of the operator $\bar q\Gamma^X q\bar q\Gamma^Y q$
over the valence quarks
by the expectation values of the products of the
constituent quark operators $Q$
 averaged over the renormalized constituent
quark PCQM states.
  Due to the normal
ordering of the operators they should act on two different quarks.
The expectation value is proportional to the probability to find
the two constituent quarks at the same space point
$$
\int |\psi(r)|^4 d^3r \sim \frac1{4\pi R^3}
$$
 with $R$ standing for the size of the system of the three
constituent quarks
 (quark core radius), while $\psi$ is the constituent quark
wave function. The factor $\frac{1}{4\pi}$ comes
from the four
angular wave functions $\frac{1}{(4\pi)^{1/2}}$ integrated over
the solid angle.
Since the PCQM deals with a quark wave function $\psi(r)$ provided
in  explicit form, the contributions $C^{XY,f_1f_2}$ are also evaluated
explicitly.
The effect of the wave function renormalization
induced  by the interaction of the constituent quarks with
the pion provides noticable corrections in the case of scalar and
vector  structures only.

The averaging
of the operators $\bar q\Gamma^X q\bar q\Gamma^Y q$ over the sea
quarks is treated
as the expectation value of these operators in pions. The
distribution of the pion field is assumed to be the one
determined by the PCQM.

The pion expectation values where expressed in paper
\cite{x1} by using the current algebra technique
through the four-quark expectation values in vacuum. We obtain the
expressions for $P^{XY,f_1f_2}$ through these expectation values.
However, to obtain the specific numbers, we use the factorization
approximation for the vacuum expectation values, suggested first by
Shifman et al. \cite{4}. In the factorization approximation it is
  assumed that the vacuum states dominate in the sum over the
intermediate states. While there are indications that this
approximation may be violated in some of the channels \cite{s}, the
factorization was advocated recently in \cite{s2}.  Under this
approximation the expectation values $P^{XY,f_1f_2}$ are expressed
by the $\bar qq$ vacuum expectation values which are known to be
\cite{2}
\begin{equation} \label{5n}
\langle 0|\bar qq|0\rangle =-
\frac{M_\pi^2 F_\pi^2}{m_u+m_d}
\end{equation}
for each of the light flavors with $M_\pi$ and $F_\pi$
 being the mass and the decay constant of the pion, while
$m_{u(d)}$ are the current quark masses.  Here we adopt the notations
accepted in chiral perturbation theory \cite{22}.
In Eq.(\ref{5n}) $q$ stands for the $u$ or $d$ quark field and isotopic
invariance of the vacuum is assumed. The value $\langle 0|\bar
qq|0\rangle$ gives the characteristic size of the contribution of the
pion sea.  Thus the contribution of the constituent quarks
is expected to be smaller than that of the sea quarks since
$\frac1{4\pi} |\langle 0|\bar qq|0\rangle R^3| \simeq \frac15$.

In the "interference term" one
of the operators $\bar q\Gamma^X q$ acts on the
valence quarks
 while another one $\bar q \Gamma^Y q$ acts on the sea quarks.
Following our  strategy,
 we approximate the corresponding matrix elements by
those averaged over the
constituent quarks and over the PCQM pion field.
There are several possibilities
to insert this four-quark operator.
The operator can connect the pion with any of the
 constituent quarks of the nucleon.
This contribution  (the "contact interference")
is proportional to $\langle\pi|\bar q\Gamma^Y q|\pi\rangle$
which does not vanish for the scalar
case only.
Thus, only the contributions $J^{SV} = J^{VS}$ and $J^{SS}$ have
nonzero values which  are proportional to the expectation value \cite
{22}
\begin{equation} \label{5m}
\langle \pi|\bar qq|\pi\rangle = \frac{M_\pi^2}{m_u+m_d} =
- \frac{\langle 0|\bar qq|0\rangle}{F_\pi^2}.
\end{equation}
However, there is an additional small factor besides the
characteristic parameter $\langle 0|\bar qq|0\rangle $.  This factor
 reflects the small probability for the pion and the
 constituent quark
to overlap at the same space point  for a nucleon Fock state
described by the
valence quarks and a pion.
The interference process can
determine a $QQ\pi$ vertex as well,
since the pseudovector
 and pseudoscalar currents connect the pion and vacuum states.
Thus, one can consider the self-energy
diagram with one of the PCQM vertices being
replaced by the four-quark operator.
 That would be the first order diagram in the
PCQM $\pi Q$ interaction.
This "vertex interference" leads to a numerically larger contribution,
except for the case of the scalar-vector condensate.  The neutral pions
provide the contribution to the terms $J^{A,ff}$ and
$J^{Ps,ff}$ of the quarks with the
same flavor. The charged pions contribute to the condensates
$\bar u\Gamma^Ad \bar d\Gamma^A u$ and
$\bar u\Gamma^{Ps}d \bar d\Gamma^{Ps} u$.
 Thus, they provide the contributions to
all the structures $J^{X,ud}$ with  coefficients defined by the
Fierz transform.

Following the general strategy of the PCQM we include only the lowest order
contributions in the $\pi Q$ interactions.
 We also assume that only the ground states of
the constituent quarks  are included
as  intermediate states in the
self-energy diagrams.
This means that the nucleon and delta-isobars only are
included as  intermediate states of
 the nucleon self-energy. This is a
standard assumption of PCQM calculations \cite{l1,25}.

It was shown in \cite{x1} that in the scalar case in the four-quark pion
expectation value a "disconnected term",
in which one of $\bar qq$ pairs acts on
vacuum, can be singled-out in a natural way.
This is strongly pronounced in the
case of the color-singlet four-quark operator
$\bar q^aq^a\bar q^bq^b$ for which
\begin{equation}\label{5t}
\langle \pi|\bar qq\bar qq|\pi\rangle = 2\langle 0|\bar qq|0\rangle
\langle\pi|\bar qq|\pi\rangle +
\langle\pi|(\bar qq\bar qq)_{int} |\pi\rangle .
\end{equation}
The "internal" contribution
presented by the second term on the right hand
side (rhs) of Eq.(\ref{5t}) appeared to be
about an order of magnitude smaller
then the "disconnected" one, presented by
 the first term. For the color
asymmetric operator determined by
Eq.(\ref{2n}) the "disconnected" terms still
provide about $\frac34$ of the total pion expectation value.
 This leads to the
natural presentation.
\begin{equation} \label{6n}
P^{S,f_1f_2} = P^{S,f_1f_2}_{dis} + P^{S,f_1f_2}_{int}
\end{equation}
with
\begin{equation} \label{7n}
 P^{S,ud}_{dis} = \frac23\cdot 2\langle 0|\bar qq|0\rangle
 \langle N|(\bar qq)_{sea}|N\rangle
\end{equation}
for the different flavors, where $|N\rangle$ represents the
fully dressed nucleon state.
The factor $\frac23$ (it is $\frac56$  for identical flavors)
on the right hand side  of
Eq.(\ref{7n}) is the weight of  the colorless combination $\bar
q^aq^a$ in the color asymmetric expectation value defined by
Eq.(\ref{2n}).
In the case of the scalar condensate the sea quarks provide the main
contribution. A large part of it is determined by the disconnected
term, related by Eq.(\ref{7n}) to the sea-quark
contribution to the $\pi N$ sigma
term.
The internal
contributions, coming mostly from the sea quarks are
several times smaller.  The expectation values
of the operators with the same flavor $(\bar u\Gamma u)^2$ in the
other channels are also determined
 mostly by the sea-quarks. For the
mixed-flavor condensate $\bar u\Gamma u\bar d\Gamma d$ the sea
quark terms and the interference terms provide  contributions of
the same  magnitude in most of the channels.  The  valence
quarks provide a smaller correction. In the case of the
 scalar-vector condensate $P^{SV}= 0$, and for the  neutron the
 valence quarks provide the main contribution,
 while in the proton the
 interference effects contribute of the same order.

In the calculations carried out below, we use the values $F_\pi =$
92.4 MeV for the pion decay constant and the value $m_u+m_d = 11$
MeV for the sum of the light quark masses.
 Latter  value, given
in \cite{gl}, leads to the
conventional value $\langle 0|\bar qq|0\rangle =
(-245$ MeV$)^3$ at the
normalization scale of $1$ GeV. This set of values was also used in the
paper \cite{44}. Note that in the papers \cite{l1,25}
another value for the
sum of the quark masses has been used, e.g. $m_u+m_d = 14 $MeV.
This value was also given in \cite{gl} as one of
the possible ones. Both
values for $m_u+m_d$ are consistent with nowadays experimental
data \cite{dpg}.

We present the results for the condensates $(\bar u\Gamma^X u)^2$
both for proton and neutron. The values of the $(\bar d\Gamma^X d)^2$
condensates are determined by the isotopic invariance  relations

$$
\langle p|(\bar d\Gamma^X d)^2|p\rangle =
\langle n|(\bar u\Gamma^X u)^2|n\rangle ,
$$
$$
\langle n|(\bar d\Gamma^X d)^2|n\rangle =
\langle p|(\bar u\Gamma^X u)^2|p\rangle
$$
while for  the mixed flavor condensates we have
$$
\langle p|\bar u\Gamma^X u\cdot \bar d\Gamma^X d|p\rangle =
\langle n|\bar u\Gamma^X u\cdot \bar d\Gamma^X d|n\rangle
$$
except for the pseudoscalar case $\Gamma^X=\Gamma^Y=\gamma_5$. In the
pseudoscalar case an explicit dependence of the interference terms on
the current quark masses $m_{u,d}$ causes  contributions which
break the isotopic invariance. These terms are numerically small.

The results enable to obtain also values
for the condensates $\bar
u\Gamma^X d\cdot \bar d\Gamma^X u$. This can be done  by using the
Fierz transform.

We compare the value of the  contribution $\langle N|(\bar
uu+\bar dd)^2|N\rangle$ with the value obtained in \cite{44} in the
framework of the NJL model under certain additional assumptions.
The values appear to differ by about 70$\%$.

For the sake of simplicity we shall use the wording "scalar",
"pseudoscalar", etc. condensates for the  expectation values of
the operators with the repeated Lorentz structures $\bar q\Gamma
q\bar q\Gamma q$. Thus the "scalar" expectation values are rather
"scalar-scalar" ones, etc.

\section{Contribution of the valence quarks}
In this Section we calculate the contribution
to the four-quark expectation
values arising from  averaging over the system
of three valence quarks. Using the results of Appendix A we present
this contribution by
\begin{equation} \label{11n}
C^{XY,f_1f_2} =
 \langle \phi_0|\int d^3x q^{f_1}(x)\Gamma^X q^{f_1}(x)
 \bar q^{f_2}(x)\Gamma^Y q^{f_2}(x)
 |\phi_0\rangle
\end{equation}
with $\phi_0$ denoting the nucleon
 as a bound state of three valence quarks.
Our main assumption here is that the matrix element in the rhs of
Eq.(\ref{11n}) is approximated by the matrix
element of the renormalized
constituent quark  operators $Q^r$, i.e.
\begin{equation} \label{11o}
C^{XY,f_1f_2} =
 \langle \phi_0|\int d^3x \bar Q^{r,f_1}(x)\Gamma^X Q^{r,f_1}(x)
 \bar Q^{r,f_2}(x)\Gamma^Y Q^{r,f_2}(x)
 |\phi_0\rangle
\end{equation}

The renormalization effects are expected to manifest themselves
through small corrections only. As we shall see below, these
corrections are of
the order of several percent only, except for the scalar and vector
structures. Thus, we
start with the unrenormalized constituent  quark operators
 $Q$ in which the pion cloud is not
included.
The corresponding contribution

\begin{equation} \label{11nt}
\stackrel{o}{C}\,\hspace*{-.2cm} ^{XY,f_1f_2} =
 \langle \phi_0|\int d^3x  \bar Q^{f_1}(x)\Gamma^X Q^{f_1}(x)
 \bar Q^{f_2}(x)\Gamma^Y Q^{f_2}(x)
 |\phi_0\rangle
\end{equation}
is illustrated by Fig.1.
 The constituent quark
operators provide a nonzero value
while acting on different quarks
of the $\phi_0$ system only. This is due to
their normal ordering. Thus
we find immediately

\begin{equation} \label{12n}
\stackrel{o}{C}\,\hspace*{-.2cm} ^{XY,uu}_n=0
\end{equation}
for the neutron. Using Eq.(\ref{11nt}) we obtain expressions for the
contributions of the constituent quarks through the wave functions
$\psi_i(x)$. Assuming that  the constituent quarks $U$ and $D$ are
described by the same wave functions $\psi_u(x) = \psi_d(x) =
\psi(x)$ we present the general expressions for the proton as
\begin{equation} \label{13n}
\stackrel{o}{C}\,\hspace*{-.2cm} ^{X,uu}_p=\int d^3x {\cal F}(x)
\end{equation}
with
\begin{equation} \label{14n}
{\cal F}(x) = \hat P \bar\psi(x)\Gamma^X\psi(x)
\cdot \bar\psi(x)\Gamma^X\psi(x) .
\end{equation}
Here $\hat P$ stands for the projection on the symmetric spin state
of the two-quark system, while the total antisymmetrization is
provided by the color variables. For the condensate $\bar
u\Gamma^Xu\bar d\Gamma^Xd$  we find  both for proton and neutron
\begin{equation} \label{15n}
\stackrel{o}{C}\,\hspace*{-.2cm} ^{X,ud} = 2\int d^3x {\cal F}(x)
\end{equation}
since there are two $u d$ pairs.

The invariant coefficients of the rhs of
Eqs.(\ref{1m}) and (\ref{1l}) can
be obtained in a specific reference frame.
Assuming $i$ and $j$ to be
the three-dimensional indices, corresponding to the four-dimensional
indices $\mu$ and $\nu$, we find in the rest frame of the nucleon
\begin{equation} \label{10}
a_c = -\frac 13 C^{XX,f_1f_1}_{ij}\delta_{ij}, \quad b_c =
C^{XX,f_1f_1}_{00} - a_c
\end{equation}
for the coefficients of Eq.(\ref{1m}), i.e.  for the
vector and pseudovector cases.  Denoting
the three-dimensional indices
 corresponding to the four-dimensional indices
  $\alpha$ and $\beta$ as
$k$ and $l$, respectively, we obtain in the same frame
\begin{equation} \label{11}
a_c^T = \frac16
C^{T,uu}_{ijkl}\delta_{ik}\delta_{jl},
\quad b_c^T = -a_c^T -\frac 13
\delta_{jl}C^{T,uu}_{0j,0l}.
\end{equation}

Above equations  are true for any constituent quark model.
In the specific case of the PCQM the wave functions of both $U$ and $D$
constituent quarks are  \cite{25}
\begin{eqnarray} \label{26}
\psi(\vec x) = N e^{(-\frac{\vec x^2}{2R^2})}
\left(
\begin{array}{c}
\chi\\
i\beta\frac{(\vec\sigma \vec x)}{R}\chi\\
\end{array}
\right)
\end{eqnarray}
with the normalization constant
\begin{equation} \label{27}
N=[\pi^{3/2}R^3(1 + \frac32 \beta^2)]^{-1/2}
\end{equation}
and $\chi$ being the two-component spinor.

The model parameters
\begin{equation} \label{28}
\beta = 0.39, \quad R=(0.6\pm 0.05) \mbox{fm}
\end{equation}
are fitted to reproduce the value of the axial coupling constant and
 of the proton charge radius. We will present the numerical
values for the mean value of $R$=0.6 fm.

A straightforward calculation provides
 for the expectation values
  $\stackrel{o}{C}\,\hspace*{-.2cm} ^{XY,uu}_p$  in the proton
\begin{equation} \label{34n}
\stackrel{o}{C}\,\hspace*{-.2cm} ^{S,uu}_p
= (1 - \frac32\beta^2 +\frac{15}{16}\beta^4){\cal N}^2
\end{equation}
with
\begin{equation} \label{35n}
{\cal N}^2 = \frac{N^2}{2^{3/2}(1 + \frac32\beta^2)}
\end{equation}
while the value is zero for neutron. For the
$\bar uu\bar dd$ condensate we get
\begin{equation} \label{36n}
\stackrel{o}{C}\,\hspace*{-.2cm} ^{S,ud} =
2(1 - \frac32\beta^2 +\frac{15}{16}\beta^4){\cal N}^2 .
\end{equation}
The details of the calculations for the other structures are
presented in the Appendix B.

For the structures $(\bar u\Gamma^Xu)^2$ in the proton we find
for the pseudoscalar case
\begin{equation} \label{37n}
\stackrel{o}{C}\,\hspace*{-.2cm} ^{Ps,uu}_p = -\beta^2{\cal N}^2,
\end{equation}
while in the vector channel
\begin{equation} \label{38n}
\stackrel{o}{a}\,\hspace*{-.2cm} ^V_{C,p} = -\frac23\beta^2{\cal N}^2 ,
\quad \stackrel{o}{b}\,\hspace*{-.2cm} ^V_{C,p} =
(1+\frac{13}{6}\beta^2 + \frac{15}{16}\beta^4){\cal N}^2,
\end{equation}
and for the pseudovector case
\begin{equation} \label{39n}
\stackrel{o}{a}\,\hspace*{-.2cm} ^A_{C,p} =
-\frac13(1-\frac12\beta^2 + \frac{15}{16}\beta^4){\cal
N}^2 ,
\quad \stackrel{o}{b}\,\hspace*{-.2cm} ^A_{C,p} =
-\stackrel{o}{a}\,\hspace*{-.2cm} ^A_{C,p} .
\end{equation}
Note that
$\stackrel{o}{b}\,\hspace*{-.2cm} ^A_C =
-\stackrel{o}{a}\,\hspace*{-.2cm} ^A_C$
 since the matrix element of the time
component of the  pseudovector operator turns to zero. This is true
for the solution of the Dirac equation in any effective field.  For
the tensor case we get
\begin{equation} \label{42n}
\stackrel{o}{a}\,\hspace*{-.2cm} ^T_C =
\frac13(1+\frac12\beta^2 + \frac{15}{16}\beta^4){\cal N}^2 ,
\quad \stackrel{o}{b}\,\hspace*{-.2cm} ^T_C =
-\frac13 (1+\frac72\beta^2 + \frac{15}{16}\beta^4){\cal
N}^2 .
\end{equation}
Following the previous analysis these values turn to zero for
the neutron.

Turning to the case of  different flavors, we find the expectation
 values of the operators
$\bar u\Gamma^Xu \bar d\Gamma^Xd$ in a nucleon to be
 twice as large as the values of $(\bar u\Gamma^X u)^2$ in the proton
\begin{equation} \label{42nn}
\stackrel{o}{C}\,\hspace*{-.2cm} ^{X,ud}_p =
 2 \stackrel{o}{C}\,\hspace*{-.2cm} ^{X,uu}_p .
\end{equation}

We also present  an  example of
the condensate for the mixed scalar-vector
structure $T^{SV,du}_\mu = \bar dd\bar u\gamma_\mu u$, which is needed
in applications.  In the rest frame of the nucleon only the time
component of the vector $T_\mu$ survives, providing
\begin{equation} \label{39}
 \stackrel{o}{C}\,\hspace*{-.2cm} ^{SV,du} =
 2(1-\frac{15}{16}\beta^4){\cal N}^2.
\end{equation}

It is convenient to express the values in "units" of the value
$\varepsilon^3_0$, defined as
\begin{equation} \label{31}
\varepsilon^3_0 = -\langle 0|\bar qq|0\rangle, \quad
\varepsilon_0\simeq 245 \mbox{ MeV}.
\end{equation}

To get a feeling for the relative size of the
contributions, we present the
 numerical value
\begin{equation} \label{50}
{\cal N}^2 = 1.50\cdot 10^{-3} \mbox{GeV}^3
\simeq 0.10\varepsilon^3_0
\end{equation}
which is the result of the straightforward computation of the rhs of
Eq.(\ref{35n}).

The $\pi Q$ interactions provide the nonzero values
of the condensates $\bar u\Gamma^X u\bar u\Gamma^Y u$ in the
neutron. This happens since the four-quark operators connect the
only valence $U$ quark with the intermediate $U$ quark of
the $\pi^- U$ self-energy loop of the valence $D$ quark. The value
is
\begin{equation} \label{50a}
C^{X,uu}_n = -2\frac{\partial \Sigma^-}{\partial E}
 \stackrel{o}{C}\,\hspace*{-.2cm} ^{X,uu}_p
\end{equation}
with $\Sigma^-$ standing for the contribution of $\pi^-$ to the
self-energy of the valence quark with the energy $E$. The direct
calculation gives $\frac{\partial \Sigma^-}{\partial E} = -0.082$.

Now we take into account the change $\delta\psi$  of the shape
of the single quark wave function $\psi(x)$,
caused by renormalization \cite{l1}
\begin{equation}\label{6-1}
\psi^r(x) = \psi(x) + \delta\psi(x), \quad
\delta\psi(x) = \Lambda(h(x) + \gamma_0)\psi(x)
\end{equation}
with
$$
\Lambda = \frac{\delta m}{2} \,
\frac{\beta R}{1 + \frac32\beta^2} , \quad h(x) = \frac{\frac12 +
\frac{21}{4}\beta^2}{1+\frac32\beta^2} -
\frac{x^2}{R^2} .
$$

Here  $\delta m < 0$ is the shift of the effective mass of
the constituent quark caused by the pion cloud.
The numerical values are $\Lambda = -2.5\cdot 10^{-2}$,
$h(x) = 1.06 -\frac{x^2}{R^2}$.

The term containing the function $h(x)$ provides  corrections
which do not exceed 3$\%$. This happens due to the strong
cancellations of the two terms, composing $h(x)$. We shall neglect
these corrections. The term containing the Dirac matrix $\gamma_0$
 mixes the Lorentz structure of the condensates.
 It provides nonvanishing
contributions to the scalar, vector
 and scalar-vector expectation values.

Thus, we obtain for renormalized values defined by  Eq.(\ref{11o})
\begin{equation} \label{6-2}
C^{S,uu}_p = \stackrel{o}{C}\,\hspace*{-.2cm} ^{S,uu}_p
+ 4\Lambda \stackrel{o}{C}\,\hspace*{-.2cm}^{SV,uu}_p
\frac1{\gamma}\, , \quad
C^{S,ud} = \stackrel{o}{C}\,\hspace*{-.2cm} ^{S,ud}
+ 2\Lambda (\stackrel{o}{C}\,\hspace*{-.2cm}^{SV,du} +
\stackrel{o}{C}\,\hspace*{-.2cm}^{SV,ud})
\frac1{\gamma}\, ,
\end{equation}
$$
C^{SV,du} \approx \, \stackrel{o}{C}\,\hspace*{-.2cm} ^{SV,du}
+ 2\Lambda (\stackrel{o}{C}\,\hspace*{-.2cm}^{S,ud}\gamma +
  \stackrel{o}{b}\,\hspace*{-.2cm} ^V_C \, \frac{1}{\gamma})
$$
with $\gamma = \frac{1-\frac32\beta^2}{1 + \frac32\beta^2}$.
 Thus, the scalar-scalar condensates are
reduced by 16$\%$ due to the renormalization effects. The
scalar-vector condensate is reduced by 11$\%$. Also, in the
vector case we have
\begin{equation}\label{6-3}
a^V_C = \, \stackrel{o}{a}\,\hspace*{-.2cm} ^V_C , \quad
b^V_C = \, \stackrel{o}{b}\,\hspace*{-.2cm} ^V_C
 (1+4\Lambda\gamma)
\end{equation}
 reducing the value of $\stackrel{o}{b}\,\hspace*{-.2cm}
  ^V_C$ by about 6$\%$.
For the pseudoscalar, axial and tensor structures the
corrections are negligibly small and we put
\begin{equation}\label{6-4}
 C^{X,f_1f_2}_p =
 \stackrel{o}{C}\,\hspace*{-.2cm} ^{X,f_1f_2}_p
\end{equation}
in these cases.

\section{Contribution of the sea quarks}

Now we calculate the contribution of the sea quarks. In the PCQM the
excess of the sea quarks in nucleons over the  QCD
vacuum sea inside the nucleon volume is contained in the mesons,
coupling to the constituent quarks.  In the SU(2) version of the
model, which we assume in this paper,
only the pions are included.  In the
framework of the PCQM this contribution is contained in the
next-to-leading order of the model.
In other words, it is sufficient to
include pion exchange in the one-loop approximation.

The distribution of the pion field $\pi^\alpha(x)$  is
determined by the PCQM quark-pion interaction
\begin{equation}\label{qp}
H_I(x) = \bar\Psi(x)i\gamma_5
\frac{S(x)\tau^{\alpha}\pi^\alpha(x)}{F_\pi} \Psi(x)
\end{equation}
where $\Psi(x)$ represents the SU(2) doublet of light quarks,
while $S(x)$ is the effective scalar field.

In the one-loop approximation of the PCQM the pions
are contained in the
 constituent quark self-energy diagrams and in the diagram,
describing the pion exchange between the constituent quarks (Fig.2).
The contribution of the sea quarks (see Eq.(\ref{2})) can be
presented as
\begin{equation} \label{52}
P^{XY,f_1f_2} = \sum_\alpha\langle\pi^\alpha|T^{XY,f_1f_2}|\pi^
\alpha\rangle
(\frac{\partial\Sigma^\alpha}{\partial M_\pi^2} +
\frac{\partial\Lambda^\alpha}{\partial M_\pi^2})
\end{equation}
with $\Sigma^\alpha$ and $\Lambda^\alpha$
standing for the self-energy and exchange
contributions of the pions $\pi^\alpha$ $(\alpha =+,-,0)$.
A similar presentation was actually
used in \cite{25} for the calculation of
the sigma term. In that case $T$ was the scalar quark operator
$\bar qq$.

The rhs of Eq.(\ref{52}) can be simplified by noticing that in
the PCQM
the relation for the total energy shifts
caused by the self-energy and
 exchange  diagrams \cite{a3}
\begin{equation} \label{53}
\sum_\alpha \Lambda^\alpha = \frac{10}{9}\sum_\alpha \Sigma^\alpha
\end{equation}
holds also for each pion $\alpha$ separately,
 when limiting single quark lines to the ground state.

Using Eq.(\ref{53}) we present the total pion contribution
to the nucleon mass  as
$$
\Sigma_t = (1 + \frac{10}{9} ) \sum_\alpha \Sigma^\alpha.
$$
This leads to
\begin{equation} \label{54}
P^{XY,f_1f_2} = \frac{3\beta^0+4\beta^++2\beta^-}{9}\frac{\partial
\Sigma_t}{\partial M_\pi^2}
\end{equation}
for the proton with
\begin{equation} \label{55}
\beta^\alpha = \langle\pi^\alpha|T^{XY,f_1f_2}|\pi^\alpha\rangle .
\end{equation}
The coefficients multiplied by the
expectation values $\beta^\alpha$ are
determined by the numbers of the quarks which can emit the pion
$\pi^\alpha$ and by the strength of the $\pi QQ$ vertex.
For example, $4\beta^+$
means that there are two quarks (these are $U$-quarks) coupling to
 a $\pi^+$, and each of the $\pi^+ D U$
  vertices contributes the factor
 $\sqrt 2$, etc.

For the neutron we get
\begin{equation} \label{56}
P^{XY,f_1f_2} = \frac{3\beta^0+2\beta^+ + 4\beta^-}{9}\frac{\partial
\Sigma_t}{\partial M_\pi^2}.
\end{equation}
The value $\frac{\partial \Sigma_t}{\partial M_\pi^2}$
 was evaluated
earlier in the calculation of the sigma term \cite{25},
providing
\begin{equation} \label{57}
\frac{\partial\Sigma_t}{\partial M_\pi^2} \approx 1.3
\mbox{GeV$^{-1}$}.
\end{equation}

The pion expectation values $\beta^\alpha$ can be expressed  by
the vacuum expectation values of the four-quark operators. This was
done in \cite{x1} by using the reduction formula obtained by Lehmann,
Symanzik and Zimmerman \cite{lsz}.
 Due to the partial conservation of the
axial current (PCAC) the pion state vector
 can be expressed by
 the vacuum as (see e.g.  \cite{Yn})
\begin{equation} \label{58}
|\pi^\alpha\rangle =
\frac{1}{\sqrt 2 F_\pi M_\pi^2}\partial_ \mu{\cal A}^\alpha_ {\mu
5}(x)|0\rangle .
\end{equation}
Here  ${\cal A}^\alpha_{\mu5}(x)$ is the axial current of the light
quarks
\begin{equation} \label{59}
{\cal A}^-_{\mu5}(x) = \sum_c\bar d^c(x) \gamma_\mu\gamma_5 u^c(x)
\end{equation}
with $c$ being the color index.
(We shall assume the summation over the colors in all the equations presented
below.) This enables to present the
expectation values defined in Eq.(\ref{55}) by the vacuum
matrix elements  \cite{x1}
\begin{equation} \label{60}
\beta^\alpha = \frac{1}{F_\pi^2}\langle 0|B^\alpha |0\rangle
\end{equation}
with
\begin{equation} \label{61}
B^\alpha = \frac{1}{2V}\int d^3x dy_0 dz_0 \delta (x_0-y_0)\delta
(z_0-x_0)
[\bar Q^\alpha_5(z_0), [Q^\alpha_5(y_0),T^{XY,f_1f_2}(x)]].
\end{equation}
Here $V$ is the normalization volume
and in the double commutator occur the axial charges
$Q^\alpha_{5}$, corresponding to the
axial current ${\cal A}^\alpha_{\mu5}(x)$. For example, in the
scalar channel it was found
\begin{equation} \label{62}
B^{\pm} = -(\bar u^a u^b\bar u^{a'} u^{b'}
+ \bar u^au^b\bar d^{a'}d^{b'}
+ \bar d^a\gamma_5 u^b\bar u^{a'}\gamma_5 d^{b'})
(\delta_{ab}\delta_{a'b'} - \delta_{ab'}\delta_{a'b})
\end{equation}
for the operator $T^{S,uu}$, averaged over the charged pions
$\pi^{\pm}$.  The total contribution of the pion cloud to the $\bar
uu\bar uu$ condensate in the $SS$ channel - see Eqs.(\ref{54}),
(\ref{56}) - takes the form
\begin{equation} \label{63}
P^{S,uu} = -\frac{2}{3{F_\pi}^2}
 \langle 0|2\bar u^a u^b\bar u^{a'}u^{b'}
 + \bar u^a u^b \bar d^{a'}d^{b'}
 + \bar u^a\gamma_5 u^b \bar u^{b'}\gamma_5 u^{a'}
 + \bar u^a\gamma_5 d^b\bar d^{b'}\gamma_5 u^{a'})|0\rangle
\end{equation}
$$
 \times (\delta_{aa'}\delta_{bb'} -
  \delta_{ab'}\delta_{ba'}).
 \frac{\partial\Sigma_t} {\partial M_\pi^2} .
$$

In  \cite{x1} the pion expectation values were expressed
by the ones of the
 vacuum for all  channels. Thus similar equations can be
presented for all structures.  Since only the vacuum expectation
values are involved, we find for the coefficients in the rhs of
Eqs.(\ref{1m}), (\ref{1l})
\begin{equation} \label{63n}
b_{p,n}^{V,A,T} = 0.
\end{equation}
To avoid complicated formulas we shall
present the final results in the framework of the factorization
hypothesis for the vacuum expectation values.

It is convenient to present
\begin{equation} \label{64}
\delta_{aa'}\delta_{bb'} - \delta_{ab'}\delta_{ba'} =
\frac 23 \delta_{aa'}\delta_{bb'} - \frac12 \sum_\rho
 \lambda^\rho_{aa'}
\lambda^\rho_{bb'}
\end{equation}
with $\lambda^\rho$ standing for the SU(3) Gell-Mann matrices
normalized by the relation Sp $ \lambda^\rho\lambda^\tau =
2\delta^{\rho\tau}$.

In the factorization approximation we find for  quarks of the
same flavor
\begin{equation} \label{65}
\langle 0|\bar q\Gamma_r q\bar q\Gamma_s q|0\rangle = \frac1{16}
[\mbox{Sp}\Gamma_r\,
\mbox{Sp}\Gamma_s -\frac13 \mbox{Sp}(\Gamma_r\Gamma_s)]
(\langle 0|\bar qq|0\rangle)^2
\end{equation}
for any $4\times 4$ matrices $\Gamma_{s,r}$
 acting on Lorentz indices.
If the quarks have  different flavors we come to
\begin{equation} \label{66}
\langle 0|\bar q_i\Gamma_r q_i\bar q_j\Gamma_s q_j|0\rangle =
\frac1{16}\mbox{Sp}\Gamma_r\,\mbox{Sp}\Gamma_s \langle 0|\bar
q_iq_i|0\rangle \langle 0|\bar q_jq_j|0\rangle \end{equation} and
\begin{equation} \label{67}
\langle 0|\bar q_i\Gamma_r q_j\bar q_j\Gamma_s q_i|0\rangle =
-\frac1{48} \mbox{Sp} \Gamma_r\Gamma_s \langle 0|\bar q_iq_i|0\rangle
 \langle 0|\bar
q_jq_j|0\rangle.
\end{equation}

For the matrices $\tilde \Gamma ^\rho_{r,s} = \Gamma_{r,s}\lambda^
\rho$ this approximation provides
\begin{equation} \label{68}
\langle 0|\sum_\rho\bar q_i\tilde \Gamma_r^\rho q_j\bar q_j\tilde
\Gamma_s^\rho
q_i|0\rangle =
-\frac1{9} \mbox{Sp} (\Gamma_r\Gamma_s) \langle 0|\bar q_iq_i|0\rangle
\langle 0|\bar
q_jq_j|0\rangle
\end{equation}
which is true for $i=j$ and $i\neq j$, while
\begin{equation} \label{69}
\langle 0|\sum_\rho\bar q_i\tilde \Gamma_r^\rho q_i\bar q_j\tilde
\Gamma_s^\rho
q_j|0\rangle = 0
\end{equation}
for $i\neq j$.

In the factorization approximation the contribution of the sea
quarks contains the factor
\begin{equation} \label{69n}
\frac{(\langle 0|\bar uu|0\rangle)^2}{F_\pi^2} =
-\,\frac{M_\pi^2}{m_u + m_d}\langle 0|\bar uu|0\rangle
\end{equation}
- see Eq.(\ref{5n}), and we can present
\begin{equation} \label{70}
P^{X,f_1f_2}_{p,n} = \frac{M_\pi^2}{m_u+m_d}\frac{\partial \Sigma_t}
{\partial
M_\pi^2}\ \varepsilon^3_0 \ S^{X,f_1f_2}_{p,n}.
\end{equation}
Here we denoted $P^{XX,f_1f_2}_{p,n} = P^{X,f_1f_2}_{p,n}$ and
 the rhs of Eq.(\ref{55}) turns to zero for $X\neq Y$. The
lower index denotes proton or neutron.

Using the results of  \cite{x1}, we find for the same flavors
\begin{equation} \label{71}
S^{S,uu}_p = S^{S,uu}_n = -\frac{16}{9},
\quad  S^{Ps,uu}_p = S^{Ps,uu}_n =
-\frac89,
\end{equation}
$$
S^{V,uu}_p = -S^{A,uu}_p = -\frac29 g_{\mu\nu}, \quad
S^{V,uu}_n = -S^{A,uu}_n = -\frac29 g_{\mu\nu},
$$
$$
S^{T,uu}_p = S^{T,uu}_n = -\frac 49s_{\mu\nu,\alpha\beta}
$$
with the tensor $s_{\mu\nu,\alpha\beta}$ defined by Eq.(\ref{9}).

For the quarks of different flavors we obtain for the proton and
neutron
\begin{equation} \label{72}
S^{S,ud} = -\frac{14}{9}, \quad
 S^{Ps,ud} = \frac29, \quad S^{V,ud} = -S^{A,ud} = \frac29 g_{\mu\nu},
\quad S^{T,ud} = -\frac29 s_{\mu\nu,\alpha\beta}.
\end{equation}

We now show  that in the scalar channel the disconnected terms
are separated in a natural way. We have the result
\begin{equation} \label{73}
S^{S,uu}_{p,n} = - \frac53 - \frac19, \quad S^{S,ud}_{p,n} =
-\frac43 - \frac29.
\end{equation}

As it was shown in  \cite{x1} the expectation values of the scalar
four-quark operators are dominated by the disconnected terms with
one of the $\bar qq$ pairs coming from the vacuum. This corresponds to
the approximation
\begin{equation} \label{74}
S^{S,uu}_{p,n} = S^{uu}_{dis} = - \frac53 ,
\quad  S^{S,ud}_{p,n} = S^{ud}_{dis} = -\frac43 .
\end{equation}
On the other hand, the factor $\frac{M_\pi^2}{m_u+m_d}$ in the rhs
of Eq.(\ref{70}) is just the expectation value of the operator $\bar
qq$ in the pion - see Eq.(\ref{5m}). Then we have
\begin{equation} \label{75}
\frac{M_\pi^2}{m_u+m_d}\frac{\partial\Sigma_t}{\partial M_\pi^2} =
 \langle \pi|\bar qq|\pi\rangle \frac{\partial \Sigma_t}
 {\partial M_\pi^2} =
 \frac12 \langle N|\bar uu+\bar dd|N\rangle_{sea}.
\end{equation}
Thus, in the scalar channel there is a contribution of disconnected
terms
\begin{equation} \label{76}
 P^{S,uu}_{dis;p,n} = \frac56 \cdot 2\langle 0|\bar uu|0\rangle
\langle N|\bar uu|N\rangle_{sea}
\end{equation}
corresponding to the approximation, expressed by Eq.(\ref{74}).  Of
course, there is a similar expression for $ P^{S,ud}_{dis;p,n}$
\begin{equation} \label{76n}
 P^{S,ud}_{dis;p,n} = \frac23\left(
\langle 0|\bar uu|0\rangle\langle N|\bar dd|N\rangle_{sea} +
\langle 0|\bar dd|0\rangle\langle N|\bar uu|N\rangle_{sea}
\right).
\end{equation}

\section{Interference terms}

We now  turn to the situation when one of
$\bar q\Gamma^X q$ operators acts on the
constituent quark while the other one acts on the pion. In the
one-loop approximation of the PCQM this contribution
 corresponds to the
Feynman diagrams illustrated by Fig.3.

\subsection{Contact interference}
The four-quark condensate can connect
the pions of the nucleon self-energy loop
with the quarks composing the nucleon.
 The contribution can be presented as
\begin{equation}\label{6-5}
\langle N|\bar q\Gamma^X q\, \bar q\Gamma^Y q|N\rangle =
\sum_{Q,n,n'}\int
\frac{\langle \phi_0|H_I|\phi_n,\pi\rangle
\langle \phi_n|\bar q\Gamma^X
 q|\phi_{n'}\rangle
 \langle \pi|\bar q\Gamma^Y q|\pi\rangle
 \langle\phi_{n'},\pi|H_I|\phi_0\rangle}
 {(E_0 - k_{10} - E_n +i\varepsilon)
 (E_0 - k_{20} - E_{n'} +i\varepsilon) }
\end{equation}
$$
\times \Delta_\pi(k_1)\Delta_\pi(k_2)\frac{d^4 k_1}{(2\pi)^4 i}
\frac{d^4 k_2}{(2\pi)^4 i}
$$
where $k_1,k_2$ are the four-momenta of the pions,
 $\Delta_\pi(k) = 1/(k^2 -
 M_\pi^2 + i\varepsilon)$ is the pion propagator.
Recall that we include intermediate quark
 states with $n=n'=0$ only.
The state vectors  $|\phi_n\rangle$ compose the complete set
of the quark states with the energy $E_n$, index $0$
corresponds to the ground state and $H_I$ denotes the quark-pion
interaction (\ref{qp}).
 The summation is carried out over the quarks $Q$
which compose the nucleon.

In the quark language this means that the four-quark condensate
can connect the pions
 with the intermediate quark of the self-energy loop
or with another
quark. These contributions are shown in Figs.3a,b.
The corresponding exchange
diagrams are shown in Figs.3c,d.
These expectation values contain the matrix
elements
$\langle\pi|\bar q\Gamma^X q|\pi\rangle$ and
$\langle Q|\bar q\Gamma^Y q|Q\rangle$. The former has a
nonvanishing value in the
scalar case only. The latter matrix element survives
 in the scalar and vector cases only, for a
unpolarized nucleon. Thus, only the expectation values $J^{SS}$
 and $J^{SV}$ obtain nonzero values.

 The connections of the pion $\pi^\alpha$ with the intermediate
 state quark
 $I^{XY}_\alpha$ (shown in Fig.3a)
 and with another quark $K^{XY}_{\alpha}$ (shown in Fig.3b)
 are tied by the relation
\begin{equation} \label{4-5}
K^{XY}_\alpha = -2 I^{XY}_\alpha
\end{equation}
for a fixed quark flavor. This relation
can be obtained by comparing the
results of the integration over the pion energy
in the loops of the diagrams
shown in Figs.3a,c.

One can write
\begin{equation} \label{77}
I^{XY}_\alpha = -\langle\pi^\alpha|\bar q\Gamma^X q|\pi^\alpha\rangle
\int d^3zF^\alpha(z) \bar \psi(z) \Gamma^Y
\psi(z) F^{\bar\alpha}(z) .
\end{equation}
with
\begin{equation} \label{78}
F^\alpha(z) = \frac{1}{2F_\pi}\int d^3x\bar \Psi(x)i\gamma_5
\tau^\alpha \Psi(x) S(x)D_\pi(x-z)
\end{equation}
where $\Psi$ is the SU(2) doublet of the light quarks. In
Eq.(\ref{78})
\begin{equation} \label{79}
D_\pi(x) =
\frac{1}{4\pi}\frac{e^{-\mu x}}{x}
\end{equation}
is the three-dimensional pion propagator with $\mu = M_\pi$, while
\begin{equation} \label{80}
S(x) = M+cx^2
\end{equation}
is the scalar field with the parameters  \cite{25}
\begin{equation} \label{81}
M = \frac{1-3\beta^2}{2\beta R}, \quad c = \frac{\beta}{2 R^3} .
\end{equation}
The parameters $\beta$ and $R$ already occurred in the definition of the
valence quark wave function of Eq.(23).
Only  terms with the scalar structure $\Gamma^X$
provide a nonzero value. Also the integral in the rhs of
Eq.(\ref{77}) does not turn to zero for the scalar and vector
structures $\Gamma^Y$ only.

The total contribution of such interference terms to the expectation
values of the operators $\bar qq\cdot \bar q\Gamma^Y q$ $(Y = S,V)$
can be expressed by the contribution $I_0^{SY}$ of the
$\pi_0 U$ loop to the self-energy diagram of $U$ quark - see
Eq.(\ref{77})
\begin{equation} \label{81n}
J^{SY} =  - \frac23 \cdot (1 + \frac{10}{9}) I_0^{SY}\cdot n
\end{equation}
with the factor $(1+\frac{10}{9})$ taking into account the exchange
diagram shown in Fig.3c - Eq.(\ref{53}), while the coefficient
 $\frac23$ is the
weight of the color-asymmetric state.  The factor $n$
takes into account the
charge dependence of the $\pi QQ$ vertices and the number of the
corresponding diagrams.
We find $n=20$ for the $(\bar uu)^2$ condensate
in the proton, turning to $n=7$ for the neutron.
It is $n=27$ for  the $\bar
uu\bar dd$ condensate.  For the scalar-vector condensate $\bar
dd\bar u\gamma_\mu u$ we have $n=20$ for the proton and
 $n=7$ for the neutron.

Details of the calculation for the value $I_0^{SY}$ are given in
Appendix C. Here we present the result. By expressing the pion
matrix element by the vacuum one - see Eq.(\ref{5m}), we
obtain
\begin{equation} \label{81m}
I_0^{SY} = C_J A^Y\langle 0|\bar qq|0\rangle
\end{equation}
with
\begin{equation} \label{83}
A^Y = \int^{\infty}_{o} \frac{dt}{t^2}f^2(t)e^{-t^2}\varphi^Y(t),
\end{equation}
while
\begin{equation} \label{84}
f(t)=\frac{\sqrt\pi}{2}erf(t) - te^{-t^2} -
\frac{2\beta^2t^3}{2-\beta^2}e^{-t^2}.
\end{equation}
Here  we use the standard notation $erf(x) =
\frac{2}{\sqrt\pi}\int^{x}_{o}e^{-y^2}dy$ and
\begin{equation} \label{85}
\varphi^S(t) = 1-\beta^2t^2, \quad \varphi^V(t) =
1+\beta^2t^2,
\end{equation}
being caused by the matrix elements $\bar \psi\Gamma^S\psi = \bar
\psi\psi$ and $\bar \psi\Gamma^V\psi = \bar \psi\gamma_0\psi$.
For the coefficient $C_J$ we get
\begin{equation} \label{86} C_J =
\frac{\sqrt \pi}{(2\pi R F_\pi)^4}
\frac{(1-\frac12\beta^2)^2}{(1+\frac32\beta^2)^3}\simeq 8.4\cdot
10^{-2} .
\end{equation}

The interference terms provide for the scalar condensates
\begin{equation} \label{4-6}
J^{S,uu}_p = 6.2\cdot10^{-2}\varepsilon^3_0 , \quad
J^{S,uu}_n = 2.2\cdot10^{-2}\varepsilon^3_0 , \quad
J^{S,ud}_p = J^{S,ud}_n = 8.4\cdot10^{-2}\varepsilon^3_0 .
\end{equation}

For the operator $\bar dd\bar
u\gamma_\mu u$ we finally obtain
\begin{equation} \label{86n}
J^{SV,du}_{p,\mu} = 9.6\cdot10^{-2}\frac{p_\mu}{m}\varepsilon^3_0
\quad \mbox{and} \quad
J^{SV,du}_{n,\mu} = 3.4\cdot 10^{-2}\frac{p_\mu}{m}\varepsilon^3_0
\end{equation}
where the lower indices $p,n$ represent the proton and  neutron,
$J^{SV}_{p(n)\mu} = J^{SV}_{p(n)0} \delta_{\mu 0}$ in the nucleon
rest frame.

Note that the insertion of the four-quark operator can lead to
 a charge-exchange pion-quark interaction
between the points of emission and absorption of the pion by the
constituent quark.  This mechanism is also described by the diagram
of Figs.3a-d and provides a contribution to the expectation value
$\bar u\Gamma^X d\bar d\Gamma^Y u$.
The charge-exchange matrix element
$\langle \pi^0|\bar d\gamma_\mu u|\pi^+\rangle$ which is related
to  the vector part of  the weak decay amplitude
$\pi^+\rightarrow \pi^0 e^+ \nu_e$,
has a non-zero value.
However, the contribution is suppressed by an
 additional small factor
$m_q/M_\pi$ when compared to the expectation values
$\langle \pi^\alpha|\bar qq|\pi^\alpha\rangle$. When calculated
in the approach described in Sect.3 the matrix elements
$\langle \pi^0|\bar d\gamma_\mu u|\pi^+\rangle$ and
$\langle \pi^0|\bar d u|\pi^+\rangle$ vanish.
 Nonvanishing values
are provided by  corrections of the relative order $M_\pi$ to
the PCAC relation expressed by Eq.(\ref{58}).

\subsection{Vertex interference}

Another type of interference term, illustrated by Fig.3e,f, is
due to the PCAC relation \cite{Yn}
\begin{equation} \label{88}
\langle 0|\bar q\gamma_\rho\gamma_5\tau^{\alpha}
q|\pi^\alpha(k)\rangle = i\sqrt 2F_\pi k_\rho
\end{equation}
where the pseudovector current connects the pion and vacuum
states. The equations of motion lead to similar relations for the
matrix elements of the pseudoscalar operator between the pion and
vacuum states. In particular
\begin{equation} \label{88a}
\langle 0|\bar u\gamma_5d|\pi^-\rangle = -\frac{i\sqrt 2F_\pi q^2}
{m_u+m_d}
\end{equation}
where $q^2$ denotes the square of the four-momentum of
the pion.
If one of the matrices acting on Lorentz indices, i.e.
$\Gamma^Y$, has a pseudovector or a pseudoscalar structure, there
is a nonvanishing matrix element
\begin{equation} \label{2v83}
\langle \phi_0|\bar q\Gamma^Z\tau^\alpha q\cdot \bar q\Gamma^Y\tau^
{\alpha} q|\phi_0,\pi^\alpha\rangle =
\langle \phi_0|\bar q\Gamma^Z \tau^\alpha q|\phi_0\rangle
\langle 0|\bar q\Gamma^Y \tau^{\alpha} q|\pi^\alpha\rangle
\end{equation}
for any matrix $\Gamma^Z$ and
\begin{equation} \label{88b}
\langle \phi_0|\bar q\Gamma^Z\tau^\alpha q|\phi_0\rangle
 = \int d^3x\bar
\Psi(x)\Gamma^Z  \tau^\alpha \Psi(x).
\end{equation}
The sum over  color is carried out in both matrix elements in
the rhs of Eq.(\ref{2v83}).

Replacing the PCQM vertex  in the nucleon self-energy
by the vertex defined by Eq.(\ref{2v83}) we obtain (see Figs.3e,f)

\begin{equation} \label{88c}
\langle N|\bar q\Gamma^X\tau^\alpha q \bar q\Gamma^Y\tau^{\alpha}
q|N\rangle =\sum_{Q,n}\int\left(
\frac{\langle \phi_0|H_I|\phi_n;\pi^\alpha\rangle
\langle \phi_n|\bar q\Gamma^X
\tau^\alpha q|\phi_0\rangle
\langle \pi^\alpha|\bar q\Gamma^Y
\tau^{\alpha} q|0\rangle} {E_0 - k_0 - E_n
+i\varepsilon}\right.
\end{equation}
$$
\left.+ \frac{
\langle 0|\bar q\Gamma^X\tau^\alpha q|\pi^\alpha\rangle
\langle \phi_0|\bar q\Gamma^Y\tau^{\alpha}q|\phi_n\rangle
\langle \phi_n;\pi^\alpha|H_I|\phi_0\rangle }
 {E_0 -k_0 -E_n +i\varepsilon}\right)
\Delta_\pi(k)\frac{d^4 k}{(2\pi)^4 i} .
$$

The rhs of Eq.(\ref{88c}) does not turn to zero only when both
$\Gamma^X$ and $\Gamma^Y$ are either pseudovector or pseudoscalar
matrices. These cases must be treated separately.
We shall use the standard PCQM approach \cite{l1,a3} where the
sum over $n$ in Eq.(\ref{88c}) is restricted to the
quark ground state.

\subsubsection{Pseudovector case}

We start with the case where both matrices $\Gamma^X$ and $\Gamma^Y$ in
Eq.(\ref{88c}) are the pseudovector ones.
The manifestation of the  vertex interference in the
self-energy diagrams is described by a certain tensor
$I_{\rho\sigma}$ with  vanishing time components - see
 Appendix B, Eq.(B11).
  Thus the integration over the energy $k_0$ in the rhs
of Eq.(\ref{88c}) can be carried out in the same way as for the
self-energy PCQM diagrams \cite{a3}. We present the
contribution as
\begin{equation} \label{2v84}
I_{\rho\sigma} = \sum_\alpha I_{\rho\sigma}^\alpha ,
\end{equation}
$$
I^\alpha_{\rho\sigma} = \int d^3z \left[
F^\alpha(z) \bar\Psi(z)
\gamma_\sigma\gamma_5\tau^{\alpha}\Psi(z)\langle 0|\bar
q(z) \gamma_\rho\gamma_5 \tau^\alpha q(z)|\pi^{\alpha} \rangle
\right.
$$
$$
\left.+ F^{\alpha}(z) \bar\Psi(z)
\gamma_\sigma\gamma_5\tau^\alpha\Psi(z)
\langle \pi^{\alpha}|\bar q(z) \gamma_ \rho\gamma_5
\tau^{\alpha} q(z)|0 \rangle \right].
$$
The two terms correspond to the manifestation of the mechanism in
the two
vertices of the one-loop self-energy diagram - see Fig.3e. The
functions $F^\alpha(z)$ are determined by Eq.(\ref{78}).

Since the operator $\bar q\tau^0 q$ does not change flavor, the
neutral pions $\pi^0$ contribute to the pseudovector condensates
$(\bar u\Gamma^A u)^2$ and  $\bar u\Gamma^Au\bar  d\Gamma
^Ad$ only. The charged pions $\pi^{\pm}$ provide  contributions
to the expectation values of the operators
$\bar u\gamma_\rho\gamma_5 d\bar d \gamma_\sigma\gamma_5 u$. Thus the
charged pions give contributions to all the structures $\bar
u\Gamma^X u\bar d\Gamma^Y d$ with the weights being determined by the
Fierz transform.

The tensor structure
\begin{equation} \label{94a}
I^\alpha_{\rho\sigma} = C^\alpha_I(g_{\rho\sigma} -
 \frac{p_\rho p_\sigma}{m^2})
\end{equation}
is  determined by setting the time components to zero.
 The coefficients $C^\alpha_I = \frac13 I^\alpha_{\rho\sigma}
 g^{\rho\sigma}$
 can be expressed through the
contributions $\Sigma^\alpha_Q$ (Fig.3e) of each quark
 to the total self-energy of the nucleon.

We start with the interference in the $\pi^+UD$ vertex. Putting
$\Gamma^X = \gamma_\sigma\gamma_5$, $\Gamma^Y = \gamma_\rho\gamma_5$
in Eq.(\ref{2v83}) and projecting it on the quark states
treated in momentum space, we obtain
\begin{equation}\label{7-1}
\langle 0|\bar d\gamma_\rho\gamma_5 u|\pi^+\rangle
 \langle U|\bar u\gamma_\sigma\gamma_5 d|D\rangle g^{\rho\sigma}
= \sqrt 2 F_\pi \int \frac{d^3k'}{(2\pi)^3}
\bar \psi(\vec k')\gamma_\rho k^\rho
i\gamma_5\psi(\vec k'-\vec k)
\end{equation}
$$
 = 2\sqrt 2F_\pi \int \frac{d^3k'}{(2\pi)^3}
 \bar\psi(\vec k')i\gamma_5 S(k)\psi(\vec k'-\vec k)
$$
with $S(k)$ standing for the scalar effective field, while $\vec k'$
 denotes the momentum of the quark.
 The last equality is due to the PCQM equation of motion.
 The rhs of Eq.(\ref{7-1}) is
 $2 F_\pi^2$ times the PCQM quark-pion vertex - Eq.(\ref{qp})

Being substituted to Eq.(\ref{2v84}) for $I^{\pm 1}$, Eq.(\ref{7-1})
provides the value $C^{\pm}_I = \frac43 F_\pi^2\Sigma^{\pm}_Q$ with
$\Sigma^{\pm}_Q$ denoting the contributions of the single quarks in
the self energies $\Sigma^{\pm}$. One can also obtain  that
$C^0_I = \frac43 F_\pi^2 \Sigma^0_Q$.
To show this, note that
\begin{equation}\label{2v91}
\bar u\Gamma u = \frac12\bar q\Gamma (\tau^0 + I) q
\end{equation}
(with $I$ standing for the $2\times 2$ unit matrix) for any $4\times 4$
matrix $\Gamma$ acting on Lorentz indices. Since $\langle 0|\bar
q\Gamma I q|\pi^0\rangle = 0$, one finds
\begin{equation}\label{2v92}
\frac1{\sqrt 2}\langle 0|\bar u\gamma_\rho\gamma_5 u|\pi^0\rangle =
\frac{i\sqrt 2 F_\pi k_\rho}{2}
\end{equation}
with the further procedure as in the charged case.

All the contributions can be expressed by the value
\begin{equation} \label{2v87}
C^0_I = \frac43 F_\pi^2\Sigma^0 ,
\end{equation}
which can be presented as
\begin{equation} \label{2v88}
C^0_I = -\frac{\sqrt 2}{6\pi^2}\,
 \frac{1}{(1+\frac32\beta^2)^2}\,
\frac1{R^3}  \int^{\infty}_{0} dy\frac{y^4 (1 - \frac{\beta^2}2
(1 + y^2))^2}{y^2 + \frac{\mu^2 R^2}2} e^{-y^2} .
\end{equation}
In the chiral limit $\mu^2 = 0$ we obtain
\begin{equation} \label{2v89}
C^0_I = -\frac{\sqrt 2}{24\pi^{\frac32}} \,
\frac{1-\frac52\beta^2 + \frac{31}{16}\beta^4 }
{(1+\frac32\beta^2)^2} \, \frac{1}{R^3} = -\frac{4.7\cdot 10^{-3}}
{R^3} = -1.14\cdot 10^{-2}\varepsilon^3_0
\end{equation}
while  for $\mu =M_\pi$ we have
\begin{equation} \label{2g90}
C^0_I = -1.0 \cdot 10^{-2}\varepsilon^3_0 .
\end{equation}

This provides
\begin{equation} \label{96a}
J^{A,uu}_{\mu\nu} = \frac23\cdot n_u (1+\frac{10}{9})
\cdot
\delta_{\mu\nu}\delta_{\rho\sigma}I^0_{\rho\sigma}
\end{equation}
where the factor $(1+\frac{10}{9})$ also includes the
exchange diagram shown in Fig.3f, - Eq.(\ref{53})
while the factor $\frac23$ is the
weight of the color asymmetric state. Thus we finally have
\begin{equation} \label{2g88}
J^{A,uu}_{\mu\nu} = -1.4\cdot 10^{-2}n_u\varepsilon^3_0 (g_{\mu\nu}
- \frac{p_\mu p_\nu}{m^2}) .
\end{equation}

Turning to  condensates of  quarks with  different flavors,
we set for the contribution of the neutral pions to the pseudovector
structure
\begin{equation} \label{2g89}
J^{A,ud}_{0\mu\nu} = (-n_u-n_d) (1+\frac{10}{9})
\cdot\frac23 \cdot\delta_{\mu\rho}\delta_{\nu\sigma} I^0_{\rho\sigma}
= 4.2\cdot 10^{-2}\varepsilon^3_0
(g_{\mu\nu}-\frac{p_\mu p_\nu}{m^2}) .
\end{equation}

The charged pions provide a direct contribution to the expectation
values of the operators
$
\bar u\gamma_\rho\gamma_5 d\bar d\gamma_\sigma\gamma_5 u
=\sum_{c,g}\bar u^c\gamma_\rho\gamma_5 d^c\bar d^g \gamma_\sigma
\gamma_5 u^g
$
with $c$ and $g$ standing for the color indices. Their contribution
to the expectation values of the operators
$\bar u^a\Gamma^X u^{a'} \bar d^b\Gamma^Y d^{b'}
(\delta_{aa'}\delta_{bb'} - \delta_{ab'}\delta_{ba'})$,
which we are looking for, is determined by the Fierz transform
\begin{equation} \label{10v1}
 u^{a'}_\alpha \bar d^b_\beta =
-\frac1{12} \sum_{A} \Gamma^A_{\alpha\beta}\ \bar
d\Gamma_A u\ \delta_{a'b}
-\frac1{64} \sum_{A\kappa} \Gamma^A_{\alpha\beta}
\lambda^\kappa_{a'b}\ \bar d\Gamma_A \lambda^\kappa u .
\end{equation}
In our case the pseudovector term with the diagonal color structure
contributes only to
\begin{equation} \label{2v90}
\bar u\Gamma^X u\bar d\Gamma^Y d = -\frac14 \cdot\frac23\cdot
\bar d\gamma_\rho\gamma_5 u\ \bar u\Gamma^X\gamma^\rho\gamma_5
\Gamma^Y d + ...
\end{equation}
Here the summation over  colors is carried out, providing the
factor $-\frac23$.  The dots denote the terms which do not contribute.

Following the previous analysis we must separate the pseudovector
component $\gamma_\sigma\gamma_5$ in the operator
$\Gamma^X\gamma^\rho\gamma_5 \Gamma^Y$ in the rhs
of Eq.(\ref{2v90}). The interference terms can be expressed by
the tensor
\begin{equation} \label{2g91}
I^C_{\rho\sigma} = 2(1 +
\frac{10}{9})\cdot(n_u+n_d)I^0_{\rho\sigma} .
\end{equation}

The tensor $I^C_{\rho\sigma}$ is obtained by the summation of the
rhs of Eq.(\ref{2v84}) over the charged pion states and over the
constituent quarks of the nucleon and by inclusion of the exchange
terms. The coefficient $C^0_I$ is given by Eq.(\ref{2v87}). The
contributions are
\begin{equation} \label{2g92}
J^{S,ud} = -\frac16 g_{\rho\sigma} I^C_{\rho\sigma} , \quad
J^{Ps,ud} = - J^{S,ud} ,
\end{equation}
$$
J^{V,ud}_{\mu\nu} =
\frac13 \delta_{\mu\rho}\delta_{\nu\sigma} I^C_{\rho\sigma}
 - \frac16 g_{\mu\nu}g_{\rho\sigma}I^C_{\rho\sigma} ,
$$
$$
J^{A,ud}_{\mu\nu} = J^{A,ud}_{0\mu\nu}
+\frac13 \delta_{\mu\rho}\delta_{\nu\sigma} I^C_{\rho\sigma}
 - \frac16 g_{\mu\nu}g_{\rho\sigma}I^C_{\rho\sigma} ,
$$
$$
J^{T,ud}_{\mu\nu,\alpha\beta} = -\frac1{24} \mbox{Sp}
(\sigma_{\mu\nu} \gamma_\rho\sigma_{\alpha\beta} \gamma_\sigma)
I^C_{\rho\sigma} .  $$

\subsubsection{Pseudoscalar case}

Now we consider  the pseudoscalar case, i.e. $\Gamma^X = \Gamma^Y
=\gamma_5$ in Eq.(\ref{88c}). If the charged pions are exchanged
the matrix elements of the quark operators between the vacuum and
the pion states are given by Eq.(\ref{88a}) which respects the
isospin symmetry. However, the contribution of the neutral pion
exchange contains the matrix elements
\begin{equation} \label{200}
\langle 0|\bar u\gamma_5 u|\pi^0\rangle =
-\frac{i F_\pi q^2}{2m_u}, \quad
\langle 0|\bar d\gamma_5 d|\pi^0\rangle =
\frac{i F_\pi q^2}{2m_d}
\end{equation}
which depend on the quark masses $m_{u,d}$ separately.
 This breaks for  $m_u\neq m_d$ the isospin symmetry explicitly.

After the integration over $k_0$ in the rhs of Eq.(\ref{88c}) (see
Appendix D) we can present the contribution  in a form similar to
the pseudovector case -Eq.(\ref{2v84})
\begin{equation} \label{201}
\tilde I = \sum_\alpha \tilde I^\alpha
\end{equation}
$$
\tilde I^\alpha = \int d^3z [F^\alpha(z) \bar \psi(z)
\gamma_5\tau^{\alpha} \psi(z) \langle 0|\bar q(z) \gamma_5
\tau^\alpha q(z) |\pi^{\alpha}\rangle
$$
$$
+ F^{\alpha}(z)\bar \psi(z)\gamma_5 \psi(z) \langle\pi^\alpha
|\bar q(z)\gamma_5 \tau^\alpha q(z)|0\rangle ]
$$
where we must put $q^2=m^2_\pi$ in the matrix elements determined
by Eqs.(\ref{88a}), (\ref{201}). The expectation values can be
expressed by the  term  $\tilde I^+$ of
Eq.(\ref{201}) corresponding to the $\pi^+$ meson
\begin{equation}\label{202}
 \tilde I^+ =  \frac{2M^2_\pi}{m_u+m_d} \int d^3x d^3y
S(x)\bar \psi(x)\gamma_5 \psi(x)
 D_\pi(x-y) \bar \psi(y)\gamma_5 \psi(y)
\end{equation}
with  $D_\pi$ and $S$ defined by Eqs.  (\ref{79}),
(\ref{80}). Numerically we get
\begin{equation} \label{203}
\tilde I^+= 7.0\cdot 10^{-2} \varepsilon^3_0 .
\end{equation}

Proceeding in the same way as in the pseudovector case we find for the
contributions of the interference terms containing the neutral $\pi^0$
mesons to $(\bar u\gamma_5 u)^2$ and $\bar u\gamma_5 u \bar d\gamma_5
d$ condensates
\begin{equation} \label{204}
J^{Ps,uu}_0 = \frac 23 (1+\frac{10}{9}) n_u \tilde I_u
\end{equation}
and
\begin{equation} \label{205}
J^{Ps,ud}_0 = -\frac 23 (1+\frac{10}{9})(n_u
\tilde I_d + n_d \tilde I_u)
\end{equation}
with
\begin{equation} \label{206}
\tilde I_{u,d} = \frac{\tilde I^+\cdot (m_u+m_d)}{4 m_{u,d}} .
\end{equation}
Using $m_u=$ 4 MeV, $m_d=$ 7 MeV we find
$$
\tilde I_u= 4.8\cdot 10^{-2}\varepsilon^3_0, \quad
\tilde  I_d= 2.8\cdot 10^{-2}\varepsilon^3_0.
$$
The charged $\pi^{\pm}$ mesons contribute to
 the expectation values
of the operators $\bar u\gamma_5 d \bar d\gamma_5u$ providing thus
the contributions to all basic structures defined by Eq.(\ref{gg}).
They contain the factor
\begin{equation} \label{207}
\tilde I_C =
 (1+\frac{10}{9}) (n_u+n_d) \tilde I^+ = 0.44 \varepsilon^3_0
\end{equation}
being
\begin{equation} \label{208}
J^{S,ud} =\frac 16\tilde I_C, \quad J^{Ps,ud}
 = J^{S,ud} + J^{Ps,ud}_0,
\end{equation}
$$
J^{V,ud}_{\mu\nu} = -\frac16 g_{\mu\nu} \tilde I_C, \quad
J^{A,ud}_{\mu\nu} = \frac16 g_{\mu\nu} \tilde I_C,
$$
$$
J^{T,ud}_{\mu\nu,\alpha\beta} = \frac 23 s_{\mu\nu,\alpha\beta}
\tilde I_C.
$$
In more sophisticated models of the pions \cite{md} the quarks
obtain large effective masses. Thus the terms $I^{\alpha}$
 will become much smaller.

\subsubsection{Mixed case}

If one of the matrices in the rhs of Eq.(\ref{2v83}) is a
pseudoscalar $(\gamma_5)$ while the other one describes the
pseudovector $(\gamma_\rho\gamma_5)$ we find contributions to the
condensates with mixed Dirac structures $\bar q\Gamma^X q \bar
q\Gamma^Y q$. If $\Gamma^X=\gamma_\rho\gamma_5, \Gamma^Y=\gamma_5$
the expectation value turns to zero when we neglect the possible
intermediate state
excitations of the constituent quarks - see Eq.(B11). The
terms with $\Gamma^Y=\gamma_\rho\gamma_5, \Gamma^X=\gamma_5$
provide  nonzero values. When we focus on the expectation
value $\bar dd\bar u\gamma_\mu u$
(which is $\bar dd\bar u\gamma_0 u$ in the
rest frame of the nucleon) among the mixed condensates, we must
calculate the expectation value of the operators $\bar u\gamma_5
d\bar d\gamma_0\gamma_5 u$ and
$\bar u\gamma_0\gamma_5 d\bar d\gamma_5
u$. Contrary to the pseudoscalar case $\Gamma^X=\Gamma^Y=\gamma_5$,
such terms do not contain the large
factor $M_\pi/(m_u+m_d)\approx 12$
- see Eq.(\ref{202}), providing thus a minor contribution $\sim
10^{-3} \varepsilon^3_0$.

\subsection{Total contribution of the interference}

Now we can present  the total contribution of the interference
terms. For the quarks of the same flavor they are presented in
Eqs.(\ref{4-6}), (\ref{2g88}), (\ref{204}) and
\begin{equation} \label{4-7}
J^{S,uu}_p = 0.06  \varepsilon^3_0 ,\quad
J^{S,uu}_n = 0.02  \varepsilon^3_0 ,
\end{equation}

\begin{equation} \label{209}
J^{Ps,uu} = 0.07 n_u \varepsilon^3_0 ,\quad
J^{A,uu}_{\mu\nu} = -0.014 n_u (g_{\mu\nu}
- \frac{p_\mu p_\nu}{m^2})\varepsilon^3_0
\end{equation}
turning to zero for the other structures. Recall that $n_u$ stands
for a number of $u$ valence quarks in a nucleon.
 For the quarks of
different flavors Eqs.(\ref{4-6}), (\ref{2g92}), (\ref{208})
provide for the proton
\begin{equation} \label{210}
J^{S,ud} = 0.22 \varepsilon^3_0 , \quad J^{Ps,ud} = -0.28
\varepsilon^3_0,
\end{equation}
$$ J^{V,ud}_{\mu\nu} = (-0.05 g_{\mu\nu}+
0.04 \frac{p_\mu p_\nu} {m^2})\varepsilon^3_0,
\quad J^{A,ud}_{\mu\nu} = 0.14 g_{\mu\nu}\varepsilon^3_0
$$
$$
J^{T,ud}_{\mu\nu,\alpha\beta} = (0.25\, s_{\mu\nu,\alpha\beta}
- 0.08\, t_{\mu\nu,\alpha\beta}) \varepsilon^3_0.
$$

Of course, the values presented by Eq.(\ref{210}) coincide for the
proton and neutron except for the pseudoscalar case where
\begin{equation} \label{211}
J^{Ps,ud}_p - J^{Ps,ud}_n = 0.03 \varepsilon^3_0.
\end{equation}
Also the value for the $(\bar u\gamma_5 u)^2$ condensate for the
proton differs from the value $(\bar d \gamma_5 d)^2$ for the
neutron by
\begin{equation} \label{212}
(J^{Ps,uu}_0)_p - (J^{Ps,dd}_0)_n = 0.06 \varepsilon^3_0.
\end{equation}
These characteristics obtain nonzero values due to the
explicit dependence on the current quark masses.
As we noted earlier, these effects would be much smaller
 if more sophisticated models for the pions are used \cite{md}.

Finally, for the scalar-vector condensates we obtain by using
Eq.(\ref{86n})
$$
J^{SV,du}_{p,\mu} = 9.6\cdot10^{-2}\frac{p_\mu}{m}\varepsilon^3_0 ,
\quad
J^{SV,du}_{n,\mu} = 3.4\cdot 10^{-2}\frac{p_\mu}{m}\varepsilon^3_0 .
$$

\section{The values of the four-quark condensates}

Now we sum the partial contributions obtained in the previous
Sections. We present the results in units of the
 characteristic scale
$\varepsilon^3_0 = -\langle 0|\bar qq|0\rangle $ (see Eq.(\ref{31})).
The partial contributions to the expectation values defined by
Eq.(\ref{1n}) are due to the constituent quarks
(denoted by $C$ and shown in Fig.1), to the pion cloud
(denoted by $P$ and shown by Fig.2)
and to the interference terms (denoted by $J$ and shown in Fig.3).
This is expressed by Eq.(\ref{2}). We do not present the values of
the parameters which are negligibly small in our scale.

\noindent
{\bf A. Scalar channel.}
Recall that in the scalar case there are specific "disconnected
terms" in which one of the products $\bar qq$ acts on the
 QCD vacuum.
Such terms emerge from the contributions of the pion cloud.
Thus in the scalar case the values of the four-quark
 condensates can
be presented as the sum of the "disconnected terms" $U_{dis}$
and the "internal terms" $U_{int}$
\begin{equation} \label{3v91}
U_{p(n)} =  U_{dis;p(n)} + U_{int;p(n)}
\end{equation}
with the indices $p, n$ denoting proton or neutron.
For the other structures the "disconnected terms" vanish.

We start by presenting the results for the disconnected terms.
Using Eqs.(\ref{70}) and (\ref{74}) we obtain
\begin{equation} \label{98a}
 U^{S,uu}_{dis} = P^{S,uu}_{dis} = - 3.83\varepsilon^3_0
\end{equation}
for both proton and neutron.

Consider now the "internal" contributions. For the scalar case they
are expressed by Eqs.(\ref{34n}), (\ref{50a}), (\ref{6-2}),
(\ref{70}),  (\ref{71}), (\ref{4-7})
\begin{equation} \label{101}
 U^{S,uu}_{int;p} = C^{S,uu}_p+P^{S,uu}_{int;p}+J^{S,uu}_p
  = -0.11 \varepsilon^3_0, \quad
\end{equation}
$$
 U^{S,uu}_{int;n} =  C^{S,uu}_n+P^{S,uu}_{int;n}+J^{S,uu}_n
 = -0.23 \varepsilon^3_0,
$$
with the partial contribution
\begin{equation} \label{101n}
 C^{S,uu}_p = 0.08 \varepsilon^3_0 , \quad
 C^{S,uu}_n = 0.01 \varepsilon^3_0 , \quad
 P^{S,uu}_{int;p} = P^{S,uu}_{int;n} = -0.25 \varepsilon^3_0,
\end{equation}
$$
 J^{S,uu}_p = 0.06 \varepsilon^3_0 ,
 \quad J^{S,uu}_n = 0.02 \varepsilon^3_0 .
$$
The total values of $U^{S,uu}_{p,n}$ (\ref{3v91}) are
the sums of Eq.(\ref{98a}) and (\ref{101})
\begin{equation} \label{101ns}
 U^{S,uu}_{p} = -3.94\varepsilon^3_0, \quad
 U^{S,uu}_{n} = -4.05 \varepsilon^3_0.
\end{equation}
In this case the sea-quarks  mainly contribute,
the rest coming from the direct
action of the four-quark operator on the constituent quarks
and from the interference terms.

Considering the mixed-flavor condensates  $\bar uu\bar dd$, we
obtain for the disconnected terms presented by Eq.(\ref{76n})
\begin{equation} \label{99}
 U^{S,ud}_{dis} = P^{S,ud}_{dis}= -3.06 \varepsilon^3_0
\end{equation}
for both proton and neutron. The internal terms are
expressed by Eqs.(\ref{36n}), (\ref{6-2}), (\ref{70}),
  (\ref{73})
 and  Eq.(\ref{210}):
\begin{equation} \label{103}
C^{S,ud}  = 0.16 \varepsilon^3_0, \quad
P^{S,ud}_{int} = -0.51  \varepsilon^3_0, \quad
J^{S,ud} = 0.22 \varepsilon^3_0 ,
\end{equation}
$$
U^{S,ud}_{int} =  C^{S,ud} + P^{S,ud}_{int} + J^{S,ud}
 = -0.13 \varepsilon^3_0 ,
$$
\begin{equation} \label{103m}
 U^{S,ud}_{p} = U^{S,ud}_{n} = U^{S,ud}_{dis} + U^{S,ud}_{int}
  = -3.19 \varepsilon^3_0.
\end{equation}
There are no "disconnected terms" in the other channels, thus $U =
U_{int}$.

\noindent
{\bf B. Scalar-vector channel.}
For this case there is no contribution if all
the four  quarks belong to the sea.  They
 contribute through the interference determined by
Eq.(\ref{86n}) while the constituent quark contribution is given
by Eq.(\ref{39}), (\ref{6-2})
\begin{equation} \label{103n}
C^{SV,du} = 0.18 \varepsilon^3_0 \quad
\end{equation}
for both proton and neutron. The interference terms are (\ref{86n})
$$
J^{SV,du}_p = 0.10 \varepsilon^3_0 , \quad
 J^{SV,du}_n = 0.03 \varepsilon^3_0 .
$$

Thus Eqs.(\ref{39}), (\ref{6-2}) and (\ref{86n}) provide
for the mixed scalar-vector condensate $\bar dd\bar u\gamma_0 u$
\begin{equation} \label{104}
U^{SV,du}_p = C^{SV,du} + J^{SV,du}_p
= 0.28 \varepsilon^3_0 ,
\end{equation}
$$
U^{SV,du}_n = C^{SV,du} +J^{SV,du}_n
 = 0.21 \varepsilon^3_0 .
$$

\noindent
{\bf C. Pseudoscalar channel.}
For the pseudoscalar case we find by using
Eqs.(\ref{37n}), (\ref{6-4}),
 (\ref{70}), (\ref{71}),  (\ref{209})
\begin{equation} \label{105}
U^{Ps,uu}_p = C^{Ps,uu} +P^{Ps,ud}+J^{Ps,uu}_p =
 -1.91 \varepsilon^3_0,
\end{equation}
 $$
U^{Ps,uu}_n = C^{Ps,uu} +P^{Ps,ud}+J^{Ps,uu}_n =
-1.96 \varepsilon^3_0 .
$$
The partial values  are
$$
C^{Ps,uu}_p = -0.02 \varepsilon^3_0 , \quad
P^{Ps,uu}_p = P^{Ps,uu}_n = -2.03 \varepsilon^3_0 , $$
$$
J^{Ps,uu}_p = 0.14 \varepsilon^3_0 ,\quad
J^{Ps,uu}_n = 0.07 \varepsilon^3_0
$$
 The numerical values are determined mostly by the
contribution $P^{Ps,uu}$ of the sea quarks. In the case of the
condensate  $\bar u\gamma_5 u\bar d\gamma_5 d$ we obtain
from Eqs.(\ref{37n}), (\ref{42nn}), (\ref{6-4}), (\ref{70}),
(\ref{72}) and
(\ref{210})
\begin{equation} \label{106}
C^{Ps,ud} = -0.03 \varepsilon^3_0 , \quad
P^{Ps,ud} = 0.51 \varepsilon^3_0 ,
\end{equation}
$$
J^{Ps,ud}_p = -0.28 \varepsilon^3_0 ,\quad
J^{Ps,ud}_n = -0.31 \varepsilon^3_0
$$
composing, following Eq.(\ref{2})
\begin{equation} \label{107}
U^{Ps,ud}_{p} = C^{Ps,ud}_p + P^{Ps,ud}_p +J^{Ps,ud}_p
 = 0.20 \varepsilon^3_0 ,
\end{equation}
$$
U^{Ps,ud}_{n} = C^{Ps,ud}_n + P^{Ps,ud}_n +J^{Ps,ud}_n
 = 0.17 \varepsilon^3_0 .
$$

The difference of the values $U^{Ps,ud}_p$
and  $U^{Ps,ud}_n$ is caused by
the explicit dependence on the current quark  masses - see
Eq.(\ref{211}).

\vspace {1cm}
For the vector, axial and tensor structures
 we present the values of the
coefficients $a^{V(A,T)}$ and $b^{V(A,T)}$.
Recall that we introduced a notation where the
partial contributions of the valence and
sea quarks and of the interference
terms are denoted by  the lower indices $C, P$ and $J$
(see the text below Eq.(\ref{9})).
 The second lower index labels the
specific nucleon. Thus, $a^V_{P,p}$ denotes
the contribution of the sea quarks
to the parameter $a^V$ of the proton, etc.
The notations $a^X_{p(n)}$ and
$b^X_{p(n)}$, labeling the vector ($V$),
 axial ($A$) and tensor ($T$) cases,
are kept for the total contributions to these parameters
for the proton
(neutron). We omit the lower index if the values coincide
for both nucleons.
Note that in all the channels the sea-quarks
 do not contribute to the parameter
$b_{p,n}$, i.e.
\begin{equation} \label{108}
b^X_{P,p(n)} = 0
\end{equation}
for all composition of flavors. Recall also that
 the interference does not
contribute to the expectation values of
the operator of the same flavors in the
vector and tensor cases - see Sect.4.

\noindent
{\bf D. Vector channel.}
By using Eqs.(\ref{38n}), (\ref{6-3}),
(\ref{70}), (\ref{71}) we obtain
\begin{equation} \label{109}
a^{V,uu}_{C,p} = -0.01\varepsilon^3_0 , \quad
a^{V,uu}_{P,p} = a^{V,uu}_{P,n} = -0.51\varepsilon^3_0 ,
\end{equation}
$$
$$
$$
a^{V,uu}_p = a^{V,uu}_{C,p} + a^{V,uu}_{P,p}
= -0.52\varepsilon^3_0 ,
\quad a^{V,uu}_n =  a^{V,uu}_{P,n} = -0.51\varepsilon^3_0
$$
while
\begin{equation} \label{110}
b^{V,uu}_p = b^{V,uu}_{C,p} = 0.13\varepsilon^3_0 , \quad
b^{V,uu}_n = b^{V,uu}_{C,n} = 0.02\varepsilon^3_0 .
\end{equation}

For the mixed-flavor condensate we find (\ref{38n}),
(\ref{42nn}),(\ref{6-3}),
 (\ref{70}), (\ref{72}), (\ref{210})
\begin{equation} \label{111}
a^{V,ud}_C = -0.02\varepsilon^3_0 , \quad
a^{V,ud}_P = 0.51\varepsilon^3_0 , \quad
a^{V,ud}_J = -0.05 \varepsilon^3_0 ,
\end{equation}
$$
a^{V,ud} = a^{V,ud}_C + a^{V,ud}_P + a^{V,ud}_J
=0.44 \varepsilon^3_0 ,
$$
which are the same for the proton and neutron,
as well as the parameters
\begin{equation} \label{112}
b^{V,ud}_C = 0.25\varepsilon^3_0 , \quad
b^{V,ud}_J = 0.04 \varepsilon^3_0 ,
\end{equation}
$$
b^{V,ud} = b^{V,ud}_C + b^{V,ud}_J = 0.29 \varepsilon^3_0 .
$$

\noindent
{\bf E. Pseudovector channel.}
Here we find by using Eqs. (\ref{39n}), (\ref{6-4}),
(\ref{70}), (\ref{71}), (\ref{209})
\begin{equation}\label{113}
a^{A,uu}_{C,p} = -0.03\varepsilon^3_0 , \quad
a^{A,uu}_{P,p} = a^{A,uu}_{P,n} = 0.51\varepsilon^3_0 ,
\end{equation}
$$
a^{A,uu}_{J,p} = -0.03\varepsilon^3_0 , \quad
a^{A,uu}_{J,n} = -0.01\varepsilon^3_0 ,
$$
$$
a^{A,uu}_p = a^{A,uu}_{C,p}+a^{A,uu}_{P,p}+a^{A,uu}_{J,p}
 = 0.45\varepsilon^3_0 ,
$$
$$
a^{A,uu}_n = a^{A,uu}_{C,n}+a^{A,uu}_{P,n}+a^{A,uu}_{J,n}
= 0.50\varepsilon^3_0
$$
and
\begin{equation}\label{114}
b^{A,uu}_{C,p} = 0.03\varepsilon^3_0 , \quad
b^{A,uu}_{J,p} = 0.03\varepsilon^3_0 , \quad
b^{A,uu}_{J,n} = 0.01\varepsilon^3_0 ,
\end{equation}
$$
b^{A,uu}_p = b^{A,uu}_{C,p} + b^{A,uu}_{J,p}
= 0.06\varepsilon^3_0 ,
$$
$$
b^{A,uu}_n =  b^{A,uu}_{J,n} = 0.01\varepsilon^3_0 .
$$

For the expectation value of the operator $\bar u\Gamma^A u
\bar d\Gamma^A d$ we get with Eqs.(\ref{39n}), (\ref{42nn}),
 (\ref{6-4}), (\ref{70}),
 (\ref{71}), (\ref{210})
\begin{equation}\label{115}
a^{A,ud}_C = -0.06 \varepsilon^3_0, \quad
a^{A,ud}_P = -0.51 \varepsilon^3_0, \quad
a^{A,ud}_J =  0.14 \varepsilon^3_0,
\end{equation}
$$
a^{A,ud} = a^{A,ud}_C +a^{A,ud}_P + a^{A,ud}_J
= -0.43 \varepsilon^3_0
$$
and
\begin{equation}\label{116}
b^{A,ud} = b^{A,ud}_C  = 0.06 \varepsilon^3_0 .
\end{equation}

\noindent
{\bf F. Tensor channel.}
Using Eqs.(\ref{42n}), (\ref{6-4}),  (\ref{70}),
(\ref{71}) we obtain
\begin{equation}\label{117}
a^{T,uu}_{C,p} = 0.04 \varepsilon^3_0, \quad
a^{T,uu}_{P,p} =  a^{T,uu}_{P,n} = -1.02 \varepsilon^3_0 ,
\end{equation}
$$
a^{T,uu}_p = a^{T,uu}_{C,p} + a^{T,uu}_{P,p}
= -0.98 \varepsilon^3_0 ,
$$
$$
a^{T,uu}_n = a^{T,uu}_{C,n} + a^{T,uu}_{P,n}
=-1.02 \varepsilon^3_0
$$
while
\begin{equation}\label{118}
b^{T,uu}_p = b^{T,uu}_{C,p} = -0.05 \varepsilon^3_0.
\end{equation}

For the mixed-flavor operator (\ref{42n}), (\ref{42nn}),
 (\ref{6-4}), (\ref{70}), (\ref{72}), (\ref{210})
\begin{equation}\label{119}
a^{T,ud}_C = 0.07 \varepsilon^3_0 , \quad
a^{T,ud}_P = -0.51 \varepsilon^3_0 , \quad
a^{T,ud}_J = 0.25 \varepsilon^3_0 ,
\end{equation}
$$
a^{T,ud} = a^{T,ud}_C + a^{T,ud}_P + a^{T,ud}_J =
 -0.19 \varepsilon^3_0
$$
while

\begin{equation}\label{120}
b^{T,ud}_C = -0.10\varepsilon^3_0 , \quad
b^{T,ud}_J = -0.08\varepsilon^3_0 ,
\end{equation}
$$
b^{T,ud} = b^{T,ud}_C + b^{T,ud}_J = -0.18 \varepsilon^3_0 .
$$

\vspace{1cm}

The final results of this Section are presented
in a compact form in Tables 1,2 (keeping the values larger
0.1 in modulus).
The numbers are given in the units of $\varepsilon ^3_0
 = 1.47\cdot 10^{-2}$ GeV$^3$, see Eq.(\ref{31}).
The values $U^{X,f_1f_2}$ are
$$
U^{X,f_1f_2}_N = \langle N|:\bar q^{f_1 a}\Gamma^Xq^{f_1 a'}\cdot
\bar q^{f_2 b}\Gamma^Xq^{f_2 b'}:|N\rangle
(\delta_{a a'} \delta_{b
b'} - \delta_{a b'}\delta_{a' b})
$$
with $N=p,n$ - see Eqs.(\ref{1n}), (\ref{2n}), (\ref{1m}), (\ref{1l}).

{\bf Table 1}

\begin{tabular}{|r||r|r||r|r|}
\hline
$X$
&
$U^{X,uu}_p$
&
$U^{X,uu}_{n}$
&
$U^{X,ud}_{p}$
&
$U^{X,ud}_{n}$
\\
\hline
$S$
&
$-3.9$
&
$-4.1$
&
$-3.2$
&
$-3.2$
\\
\hline
$SV$
&
&
&
$0.3$
&
$0.3$
\\
\hline
$Ps$
&
$-1.9$
&
$-2.0$
&
$0.2$
&
$0.2$
\\
\hline
\end{tabular}

\vspace{1cm}

{\bf Table 2}

\begin{tabular}{|r||r|r|r|r||r|r|r|r|}
\hline
$X$
&
$a^{X,uu}_{p}$
&
$b^{X,uu}_{p}$
&
$a^{X,uu}_{n}$
&
$b^{X,uu}_{n}$
&
$a^{X,ud}_{p}$
&
$b^{X,ud}_{p}$
&
$a^{X,ud}_{n}$
&
$b^{X,ud}_{n}$
\\
\hline
$V$
&
$-0.5$
&
$0.1$
&
$-0.5$
&
$0$
&
$0.4$
&
$0.3$
&
$0.4$
&
$0.3$
\\
\hline
$A$
&
$0.5$
&
$0.1$
&
$0.5$
&
$0$
&
$-0.4$
&
$0.1$
&
$-0.4$
&
$0.1$
\\
\hline
$T$
&
$-1.0$
&
$-0.1$
&
$-1.0$
&
$0$
&
$-0.2$
&
$-0.2$
&
$-0.2$
&
$-0.2$
\\
\hline
\end{tabular}

\section{Summary}

We calculated the expectation values of the
four-quark QCD operators $\bar q\Gamma^X q\bar q\Gamma^Y q$
in nucleons for all basic Lorentz structures and for
compositions of the light quark flavors.

We employed previously derived results of the
 perturbative chiral quark model (PCQM) which
treats the nucleon as a system of three valence quarks
surrounded by a pion cloud. We approximate the averaging of
 the product of
operators over the valence quark by the matrix
elements of the constituent quark
operators over the PCQM  constituent quarks.
 We present the expectation values of the operators acting on
the sea quarks by the expectation values of QCD operators in pions.
 The intensity  of the pion field is determined by the PCQM
 model result.

The expectation values of the scalar and pseudoscalar operators
are Lorentz scalars. In the other channels they have a more
complicated tensor structure being determined by the
two parameters
$a^{V(A,T)}$ and $b^{V(A,T)}$ -Eqs.(\ref{1m}), (\ref{1l}). For
the quark operators with the same flavor, e.g.  $\bar q\Gamma^X
q\bar q\Gamma^X q$, the scalar and pseudoscalar condensates, as
well as the parameters $a^{V(A,T)}$ for
the other structures are dominated
by the contribution of the sea quarks.
The averaging of four $U$-quark operators over the valence
quarks in the neutron provide zero values in the lowest order of
PCQM.
This occurs because the operators $\bar uu$ should act on
different quarks while there is only one $U$-quark in the neutron.
In the case of the proton both constituent quark  and
interference terms provide minor corrections of the order of
several percent to the main contribution of the sea
quarks. In the
contrary, the sea quarks do not contribute to the coefficients
$b^{V(A,T)}$. In the vector and tensor channels the values
$b^{V,T}$ for the proton are determined by the contribution of the
constituent quarks.

In the case of the mixed-flavor condensate $\bar u\Gamma^X u \bar
d\Gamma^X d$ the role of the vertex interference increases due to
the large combinatorial factor. These terms become as important as
the sea-quark terms in most of the channels. The parameters
$b^{V(A,T)}$ are determined mostly
by the contributions of the constituent quarks.

The contributions of the sea quarks are expressed by the
expectation values of the four-quark operators in pions.
Latter values are in turn expressed by
 the expectation values of the
four-quark operators in vacuum \cite{x1}. The specific numerical
values are obtained by using the vacuum factorization
approximations \cite{4}. Thus the contribution of the sea quarks
is expressed by the well known vacuum expectation value
$\langle 0|\bar qq|0\rangle$.

In the case of the scalar-vector condensate
$\bar dd\bar u\gamma_0 u$ there is no
contribution coming from the pions only. Averaging over the
neutron is dominated
by the contribution of the constituent quarks. The
interference  and the constituent quark terms are of
the same order of magnitude in the proton.

We can draw some conclusions on the chiral properties
of the expectation values which we study in the present
paper. The contribution of the sea-quarks has the same
explicit dependence on
the pion mass $M_\pi$ as the contribution of the
sea-quarks to the expectation value
$\langle N|\bar qq|N\rangle$. The latter
 expectation value, which is proportional
 to $M_\pi^2$ times the pion-nucleon $\sigma$-term,  is
known to depend strongly on $M_\pi$. On the contrary,
our interference terms exhibit only a
weak dependence on $M_\pi$.
The valence quark contribution does not contain an
explicit dependence on $M_\pi$.

Due to the explicit dependence of the vertex
 interference terms on the
quark current masses, we have the isotopic
 symmetry breaking effects
in the pseudoscalar channel. The absolute
magnitude of this effect is numerically small with
 several units of the value
$10^{-2}\varepsilon^3_0$ for our scale $\varepsilon_0$.
The effect is much smaller if the quarks, composing pions are
assumed to have the constituent (but not current) masses \cite{md}.

In the special case of the scalar
 condensate the expectation values are
dominated by "disconnected terms" in which
one of the quark operators acts
"inside" the nucleon while the other one acts on the QCD vacuum.
This contribution comes from the sea quarks, reflecting the pion
structure \cite{x1}.

Note that a nucleon expectation value is the excess
 of the density of the quark
operator products over the vacuum density, integrated
over the volume of the nucleon
\begin{equation}\label{s-1}
\langle N|\bar q\Gamma^X q \bar q\Gamma^Y q|N\rangle =
\langle N|\int d^3x[
\bar q(x)\Gamma^X q(x) \bar q(x)\Gamma^Y q(x) -
\langle 0|\bar q\Gamma^X q \bar q\Gamma^Y q|0\rangle ]|N\rangle\ .
\end{equation}
Of course, the first term in the rhs of Eq.(\ref{s-1}) is positive.
 However, the whole rhs of Eq.(\ref{s-1}) can be negative.
  This is why some of the expectation values run negative.

In an earlier calculation \cite{44} the scalar  expectation value
$\langle N|(\bar uu +\bar dd)^2|N\rangle$ was determined
 in the framework of the Nambu--Jona-Lasinio model
\cite{1} under certain additional assumptions.
Actually, the expectation values of the color-singlet
four-quark operators $\bar q^aq^a\bar q^bq^b$
have been obtained in \cite{44}. Thus,
 to compare to the results of \cite{44} we must extend our
analysis to such operators as well.

 In the paper \cite{44} the
expectation value is presented as the composition of the
contribution of the constituent quarks $A_D$ and of $\sigma$ and
$\pi$ mesons, $A_\sigma$ and $A_\pi$.
Our contribution of the constituent quarks appears
to be several times smaller than the value of $A_D$.
The large discrepancy is not surprising,
 since the conception of the constituent
quarks is quite different in the two models.
 The meson contribution $A_\sigma + A_\pi$ of  \cite{44}
could be compared with the "internal"
sea-quark contribution of the present model,
containing the expectation value, which is
presented by the second term of the
rhs of Eq.(\ref{5t}).  The corresponding contribution $\hat P$
(with the "hat" sign labeling the color singlet operator) can be
obtained by using the formula obtained  in \cite{x1}. The result
$\hat P =\frac23\frac{\partial \Sigma_t}{\partial M_\pi^2}
\frac{\varepsilon^6_0}{F_\pi^2} $
should be compared to the sum $A_\sigma +A_\pi$ of \cite{44}.
We find $\hat P = 1.53 \varepsilon^3_0 = 2.3\cdot 10^{-2}$ GeV$^3$
while $A_\sigma +A_\pi = 3.6\cdot 10^{-2}$ GeV$^3$.
 The total values in the two models are $A_\sigma + A_\pi + A_D$
 in \cite{44} and the sum $\hat U = \hat P+\hat C +\hat J$
in our approach. We obtain $\hat C = \frac12 C;$ $\hat J = \frac32 J$
where the additional term $\hat J$ dominates in the sum
 $\hat C + \hat J$. The NJL value is $A_\sigma + A_\pi + A_D =
 6.4\cdot 10^{-2}$ GeV$^3$ while we obtain
 $\hat U = 2.5\varepsilon^3_0 =
 3.7\cdot 10^{-2}$ GeV$^3$.  The results provided by the two
 approaches differ by a factor of about 1.7.
One of the possible reasons for
 the discrepancy is that some of the
contributions have not been accounted for in both approaches.

\vspace{1.0cm}

Two of us (E.G.D. and V.A.S.) are grateful
for the hospitality of University of
T\"ubingen during their visit.
The work was  supported  by
the DFG grants: 438/RUS~113/595/0-1,
FA67/25-1,
GRK683
and by the RFBR grant 03-02-17724.

\section{Appendix A}
Here we show how the contributions to the
 four-quark expectation values,
obtained in the paper, manifest themselves
 with the help of the PCQM formalism. As an example we consider
the operator $\bar uu\bar uu$ averaged over the proton. In the
framework of the PCQM the nucleon is a system of  three constituent
quarks, where the bare three-quark state is renormalized by $\pi N$
interactions.  Thus, the physical proton state $|N\rangle$ is
expressed as
$$ |N\rangle = T exp(- i :\int^{o}_{-\infty}dt
H_I^r(t):) |\phi_0\rangle
$$
where $|\phi_0\rangle$ is the state of
three valence quarks and $H_I^r$ is the renormalized Hamiltonian of
 the  interaction between the constituent quark and the pions
which includes the counterterms.

The expectation value $\langle N|(\bar uu)^2|N\rangle$ can
then be written as
$$
\langle N|(\bar uu)^2|N\rangle =
Z^2\langle \phi_0|(\bar uu)^2|\phi_0\rangle
$$
$$
+ 2\langle \phi_0|(\bar uu)^2|\phi_0\rangle
\langle \phi_0|H_I^r|\phi_0,\pi\rangle
\langle \phi_0,\pi| H_I^r |\phi_0\rangle
$$
$$
+ \langle \phi_0|H_I^r|\phi_0,\pi\rangle
\langle \phi_0,\pi|(\bar uu)^2|\phi_0,\pi\rangle
  \langle \phi_0,\pi|H_I^r|\phi_0\rangle  .
\eqno{(A1)}
$$
Here $Z = 1 + \frac{\partial\Sigma}{\partial E}$ is
the renormalization factor, while $\Sigma$
 is the sum of the self-energies of the
constituent quarks with energy $E$.

In the next step we present each pair of the operators
 $\bar uu$ as the sum of operators acting on the
 valence  and the sea quarks
$$
\bar uu = (\bar uu)_v + (\bar uu)_s .
   \eqno{(A2)}
$$

Thus
$$
\langle \phi_0|(\bar uu)_s|\phi_0\rangle = 0,
 \quad \langle \pi|(\bar
uu)_v|\pi\rangle = 0
$$
and Eq.(A1) takes the form
$$
\langle N|(\bar uu)^2|N\rangle  =
(1 + 2\frac{\partial\Sigma}{\partial E})
\langle \phi_0|(\bar uu)_v^2|\phi_0\rangle
   \eqno{(A3)}
$$
$$
+ 2 \langle \phi_0|(\bar uu)^2|\phi_0\rangle
\langle \phi_0|H_I|\phi_0,\pi\rangle
\langle \phi_0,\pi| H_I |\phi_0\rangle
$$
$$
+ \langle \phi_0|H_I|\phi_0,\pi\rangle
\langle \phi_0|(\bar uu)_v^2|\phi_0\rangle
\langle \phi_0,\pi|H_I|\phi_0\rangle
$$
$$
+ \langle \phi_0|H_I|\phi_0,\pi\rangle
\langle \pi|(\bar uu)_s^2|\pi\rangle
\langle \phi_0,\pi|H_I|\phi_0\rangle
$$
$$
+ 2 \langle \phi_0|H_I|\phi_0,\pi\rangle
\langle \phi_0|(\bar uu)_v|\phi_0\rangle
\langle \pi|(\bar uu)_s|\pi\rangle
\langle \phi_0,\pi|H_I|\phi_0\rangle  .
$$

Since we include the quark-pion interactions
to  lowest order, the
renormalization effects are taken into account in the first term
of the rhs of Eq.(A3) only.
We put $Z^2 = 1 + 2 \frac{\partial\Sigma}{\partial E}$.

The rhs of Eq.(A3) can be simplified due to some cancellations.
The second term in the rhs describes the self-energy insertions.
These contributions are canceled by the counterterms of
the PCQM
Lagrangian \cite{l1}. Another cancellation
occurs between the third term and the part of the first term
$$
2 \frac{\partial\Sigma}{\partial E} \langle\phi_0|(\bar uu)^2_v|
\phi_0\rangle
+ \langle\phi_0|(H_I|\phi_0,\pi\rangle
 \langle\phi_0|(\bar uu)^2_v|\phi_0\rangle
 \langle\phi_0,\pi|H_I|\phi_0,\pi\rangle = 0 \, .
   \eqno{(A4)}
$$
This can be obtained in a straightforward way.
The last equation is a rather standard
cancellation of the radiative correction by the
renormalization factor. Note that the two terms in the rhs of Eq.(A4)
do not cancel totally for the operators $(\bar uu)^2$ averaged over
the neutron - see Sect.2.

Thus, Eq.(A1) takes the form
$$
\langle N|(\bar uu)^2|N\rangle =
\langle \phi_0|(\bar uu)^2_v|\phi_0\rangle
$$
$$
+ \langle \phi_0|H_I|\phi_0,\pi\rangle
\langle \pi|(\bar uu)^2_s|\pi\rangle
  \langle \phi_0,\pi|H_I|\phi_0\rangle
\eqno{(A5)}
$$
$$
+ 2\langle \phi_0|H_I|\phi_0,\pi\rangle
\langle \phi_0|(\bar uu)_v|\phi_0\rangle
\langle \pi|(\bar uu)_s|\pi\rangle
\langle \phi_0,\pi|H_I|\phi_0\rangle  \, .
$$

Now we can identify the terms in the rhs of Eq.(A5).
The first term corresponds to the contributions of the
constituent quarks.
The second
term describes the contribution of the sea quarks shown in Fig.2.
The third term presents the "interference"
 effects with one of the  $\bar uu$ pairs coming
 from pions while another one comes
from the constituent quark.
The latter can be the same as the one in the matrix
element of the interaction $H_I$ or the other one.
 These terms are shown in Fig.3a-d.
A cancellation similar to Eq.(A4) takes
place for all the operators $(\bar
u\Gamma^X u)^2$ in the proton, although in the general
 case the operator depends on the spin
variables.  However, the spin dependence
manifests itself through the operator
$(\vec \sigma^I\vec\sigma^{II})$ with $I$
and $II$ denoting the two quarks.
Since the color wave function is asymmetric,
the two quarks  compose the
spin-symmetric state being at the same space point.
 Thus, the two-quark spin
wave function $|\chi^{I,II}\rangle$ is the
eigenfunction of the operator
$(\vec \sigma^I\vec\sigma^{II})$ with
$(\vec \sigma^I\vec\sigma^{II})|\chi^{I,II}\rangle
 = |\chi^{I,II}\rangle$.
Hence, the four-quark expectation value can
be separated as a factor  and the cancellation takes
place as well as in the scalar
case.
Similar analysis can be carried out for the operators of the
general form $\bar q\Gamma^X q\bar q\Gamma^Y q$.

In the special case of the axial and pseudoscalar
operators, there can be the interference effects in the
 first order of the $\pi Q$
interaction. This happens because the matrix elements
$\langle 0|\bar q\Gamma^X q|\pi\rangle$ have nonzero
 values in these cases.
Thus, the operators $\bar q\Gamma^X q \bar q\Gamma^X q$
determine a $\pi QQ$ vertex
$\langle Q|\bar q\Gamma^X q \bar q\Gamma^X q|Q,\pi\rangle$ .
This causes the "vertex interference" contributions
$$
\langle N|\bar q\Gamma^X q \bar q\Gamma^X q|N\rangle_{intrf}
=  \langle \phi_0|H_I|\phi_0,\pi\rangle
\langle \phi_0,\pi|\bar q\Gamma^X q \bar q\Gamma^X q|\phi_0\rangle
   \eqno{(A6)}
$$
$$
+ \langle \phi_0|\bar q\Gamma^X q \bar q\Gamma^X q|\phi_0,\pi\rangle
  \langle \phi_0,\pi|H_I|\phi_0\rangle
$$
with $X$ labeling an axial or pseudoscalar. Such terms are shown
in Figs.3e,f.

In the rhs of Eq.(A6) we have
$$
\langle\phi_0,\pi|\bar q\Gamma^X q\, \bar q\Gamma^X q|\phi_0\rangle =
\langle\pi|\bar q\Gamma^X q|0\rangle
\langle\phi_0|\bar q\Gamma^X q|\phi_0\rangle
\eqno{(A7)}
$$
$$
\langle\phi_0|\bar q\Gamma^X q\, \bar q\Gamma^X q|\phi_0,\pi\rangle =
\langle 0|\bar q\Gamma^X q|\pi\rangle
\langle\phi_0|\bar q\Gamma^X q|\phi_0\rangle .
$$

We assume that the matrix elements of the
QCD operators $(\bar qq)^2_v$
over $\phi_0$ are approximated by the
matrix elements of the renormalized PCQM
constituent quark operators, i.e.
$$
\langle\phi_0|(\bar qq)^2_v|\phi_0\rangle  =
\langle\phi_0|(\bar Q^r Q^r)^2|\phi_0\rangle
\eqno{(A8)}
$$
in the first terms of the rhs of Eqs.(A3) and (A5). The renormalization
\cite{l1} means that the shape of the constituent
  quark wave function is
modified by the influence of the pion cloud.
Also, we approximate the matrix
element $\langle\phi_0|(\bar qq)_v|\phi_0\rangle =
\langle\phi_0|\bar QQ|\phi_0\rangle$ in the
third term of the rhs of Eq.(A5).

\section{Appendix B}

Except for the scalar case, the matrix
element between the two-quark
states depends on the spin orientation,
 containing the factor $(\vec
\sigma^I\vec\sigma^{II})$ with $I$ and
 $II$ denoting the two quarks.
Since the color wave function is asymmetric,
 the two quarks compose the
spin-symmetric state when being at the
same space-point. Thus we must
put
$$
\langle \chi^{I,II}|(\vec\sigma^I
 \vec\sigma^{II})|\chi^{I,II}\rangle = 1
  \eqno{(B1)}
$$
for the value of the spin operator
$(\vec\sigma^I\vec\sigma^{II})$
averaged over the spin two-quark wave function
$\chi^{I,II} $ of the
quarks $I$ and $II$.

For the scalar case we find immediately
$$
{\cal F}(x)=g(x)(1 - \beta^2 \frac{x^2}{R^2})^2
  \eqno{(B2)}
$$
with ${\cal F}(x)$ defined by Eq.(\ref{14n}),
 while $g(x)=e^{-2x^2/R^2}N^4$.
 This provides
$$
C^{S,uu}_{int} = {\cal N}^2(1-\frac32\beta^2
+\frac{15}{16}\beta^4)
  \eqno{(B3)}
$$
for the proton, with ${\cal N}$ defined by Eq.(\ref{35n}).
 For the pseudoscalar case we get
$$
{\cal F}(x) = -4\beta^2 g(x) \frac{(\vec\sigma^I\vec x)
(\vec\sigma^{II}\vec x)}
{R^2}
  \eqno{(B4)}
$$
leading to
$$
C^{Ps,uu}_{int} = -{\cal N}^2 \beta^2.
  \eqno{(B5)}
$$

For the vector and pseudovector structures
we can find in the rest
frame of the nucleon
$$
a^{V(A)}_C = -\frac13 C^{V(A)}_{ij}\delta_{ij}
  \eqno{(B6)}
$$
with $i$ and $j$ being the space indices, corresponding to
the four-dimensional indices $\mu$ and $\nu$.
 A direct calculation provides
for the vector case
$$
{\cal F}_{ij} = 4\beta^2g(x)\frac{x^2}{R^2}\cdot \frac13(\delta_{ij}
(\vec\sigma^I\vec\sigma^{II}) - \sigma^I_i\sigma^{II}_j),
  \eqno{(B7)}
$$
leading to
$$
a^V_C = {\cal N}^2(-\frac23\beta^2).
  \eqno{(B8)}
$$
To determine the coefficient $b^V_C$, we
calculate the time components
$$
{\cal F}_{00} = g(x) (1+\beta^2\frac{x^2}{R^2})^2
  \eqno{(B9)}
$$
and, since $C_{00} = a^V_C +b^V_C$, we find
$$
b^V_C = {\cal N}^2 (1+\frac{13}{6}\beta^2 + \frac{15}{16}\beta^4) .
  \eqno{(B10)}
$$

For the pseudovector case notice that the time components turn to
zero.
We introduce the notation
$$
 \kappa=
\left(
\begin{array}{c}
\chi\\
i\beta\frac{(\vec\sigma \vec x)}{R}\chi\\
\end{array}
\right)
$$

for the bispinor entering the wave function
 - Eq.(\ref{26}). We obtain
$$
\bar \kappa\gamma_0\gamma_5\kappa =0.
  \eqno{(B11)}
$$
Thus
$$
a^A_C + b^A_C = 0.
  \eqno{(B12)}
$$
As to the value of $a^A$, it can be calculated by using
Eq.(B6).  In the pseudovector case we get
$$
\bar \kappa\gamma_i\gamma_5\kappa = \sigma_i +
\beta^2\frac{(\vec\sigma
\vec x) \sigma_i(\vec\sigma \vec x)}{R^2}.
  \eqno{(B13)}
$$
By using the properties of the Pauli matrices one finds
$$
(\vec\sigma \vec x)\sigma_i(\vec\sigma\vec x)
 = 2x_i(\vec\sigma\vec
 x) - x^2\sigma_i.
  \eqno{(B14)}
$$
Thus
$$
{\cal F}_{ij}(x) = g(x)
(\sigma^I_i + \beta^2\frac{2 x_i(\vec\sigma^I
 \vec x)-\sigma^I_ix^2}
{R^2})
(\sigma^{II}_j + \beta^2\frac{2 x_j(\vec\sigma^{II}\vec x)
-\sigma^{II}_jx^2} {R^2})
  \eqno{(B15)}
$$
leading to
$$
a^A_C = {\cal N}^2(-\frac13 + \frac16\beta^2 -\frac{5}{16}\beta^4).
  \eqno{(B16)}
$$

Finally, in the tensor case we find for the space components
$$
\bar \kappa\sigma_{ij}\kappa = \varepsilon_{ijk} (\sigma_k -\beta^2
\frac{(\vec\sigma\vec x)\sigma_k(\vec\sigma\vec x)}{R^2})
  \eqno{(B17)}
$$
and the function $f$ can be  obtained by using Eq.(B14). For the
space-time components we have
$$
\bar \kappa\sigma_{0j}\kappa = -2\beta\frac{x_j}{R}
  \eqno{(B18)}
$$
and
$$
{\cal F}_{0j,0k} = \frac43\beta^2g^2(x)\frac{x^2}{R^2}\delta_{jk}
  \eqno{(B19)}
$$
with the further procedure described in the main text.

\section{Appendix C}

In order to calculate the value $I^{SY}_0$ introduced by
 Eq.(\ref{81n}), we must calculate the function $F^0(z)$ -
Eq.(\ref{78}). For the $u$-quark it takes the form
$$
F^0(z) = -\frac{N^2}{F_\pi R}\beta\int d^3x\chi^* (\vec\sigma\vec
x)\chi S(x)D_\pi(x-z)\Phi^2(x),
  \eqno{(C1)}
$$
changing the sign for the $d$-quark.
We present the pion propagator (\ref{79}) as
$$
D_\pi(x-z) = \int \frac{d^3k}{(2\pi)^3} \frac{e^
{i\vec k(\vec x-\vec z)}
e^{-i\vec x\vec a}} {k^2 + \mu^2} , \quad (a=0) .
  \eqno{(C2)}
$$
The factor $e^{-i(\vec x\vec a)} (a=0)$ is introduced in order to
simplify the calculations by expressing
$$
\vec x D_\pi(x-z) = i\vec\nabla_a D_\pi(x-z).
  \eqno{(C3)}
$$
We obtain by doing the integral over $x$
$$
 F(z) = -\frac{\pi^{3/2}}{2}\cdot \frac{N^2\beta R^4}{F_\pi} \chi^*
(\vec\sigma\vec\nabla_b) \chi (A T_1(z) + B T_2(z)), \quad (b=0)
  \eqno{(C4)}
$$
with  $A= M +\frac 52cR^2$, $B=-\frac14cR^4$,
while $c$ and $R$ are determined by Eqs.(\ref{28}),(\ref{81}) and
$$
T_1(z) = \int \frac{d^3k}{(2\pi)^3}
 \frac {e^{-i\vec k(\vec z-\vec b)
 - \frac14k^2R^2}}{k^2+\mu^2} ,
  \eqno{(C5)}
$$
$$
T_2(z) = \int \frac{d^3k}{(2\pi)^3} \frac
{e^{-i\vec k(\vec z-\vec b) - \frac14 k^2R^2}}
{k^2+\mu^2} k^2.
  \eqno{(C6)}
$$
We can evaluate the rhs of Eqs.(C5), (C6) by presenting
$$
\frac{1}{k^2+\mu^2}
 = \int^{\infty}_{o}d\alpha e^{-\alpha(k^2+\mu^2)},
  \eqno{(C7)}
$$
leading to
$$
T_1(z) =  \int^{\infty}_{o}d\alpha \int \frac{d^3k}{(2\pi)^3}
e^{i\vec k(\vec z-\vec b) -
\frac14 k^2R^2-\alpha(k^2+\mu^2)},
  \eqno{(C8)}
$$
$$
T_2(z) =  \frac{1}{\pi^{3/2}}
 \frac{1}{R^3}e^{\frac{(\vec z-\vec b)^2}
{R^2}}  - \mu^2 T_1(z).
  \eqno{(C9)}
$$

The further calculations can be simplified by assuming the chiral
limit $\mu^2=0$.
The integral in the rhs of Eq.(C8) is dominated by the values of
$z^2$ close to $\frac23 R^2$. Thus, the integral over $k^2$ is
determined by $k^2\sim \frac32\frac{1}{R^2}$, while the integral over
$\alpha$ is dominated by $\alpha\sim \frac23 R^2$. Hence, the factor
$\alpha\mu^2$ in the power of the exponent in the rhs of Eq.(C8)
is about 0.12. Since $I^{SY}_0$ provides a small correction only,
this makes the  calculation of this value in the chiral limit
$\mu^2=0$ reasonable.

Calculation of the integrals over $k$ and over $\alpha$ (by the
substitution $t=(\frac14R^2+\alpha)^{-\frac12})$ leads to
Eqs.(\ref{81m}) - (\ref{86}) of the text.

\section{Appendix D}

The integrals over the time component $k_0$ in the rhs of
Eq.(\ref{88c}) take the form
$$
X = \int\frac{dk_0}{2\pi i}\cdot
\frac{q^2}{q^2-M_\pi^2+i\varepsilon}\cdot
\frac{1}{E_0 - k_0 -E_n+i\varepsilon}
  \eqno{(D1)}
$$
with $q^2 = k_0^2 - {\vec k}^2$. We can present $X= X_1 +X_2$ with
$$
X_1 = M^2_\pi\int\frac{dk_0}{2\pi i}\cdot
\frac{1}{q^2-M_\pi^2+i\varepsilon}\cdot
\frac{1}{E_0 - k_0 -E_n+i\varepsilon} ,
  \eqno{(D2)}
$$
$$
X_2 = \int\frac{dk_0}{2\pi i}\cdot
\frac{1}{E_0 - k_0 -E_n+i\varepsilon}  .
  \eqno{(D3)}
$$
The integral $X_2$ can be expressed through the contribution of the
pole in the upper half-plane of the complex variable $k_0$. This
corresponds to the negative-energy solutions of the Dirac equation.
Such terms are neglected in framework of the PCQM. Hence, we put
$X=X_1$, leading to Eq.(\ref{202}).

\newpage

\section{Figure captions}

\noindent
Fig.1.
The contribution of the valence quarks
 to the expectation values of the four-quark operators.
  Solid lines denote the valence
quarks, the dark squares denote the four-quark operator.

\noindent
Fig.2.
The contribution of the sea quarks to
the expectation values of the
four-quark operators. The dashed lines represent the pions.
Other notations are the same as in Fig.1.
The dark circles denote the vertices of the pion-quark
interaction.

\noindent
Fig.3.
The contribution of the interference
 term to the expectation values of
the four-quark operators.
 The "contact interference" is illustrated by
Figs.3a-d. The "vertex interference" is shown in Figs.3e,f.
The permutated diagrams are not shown.
The notations are the same as in Figs.1,2.

\newpage

\begin{figure}
\vspace{-5cm}
\centering{\epsfig{figure=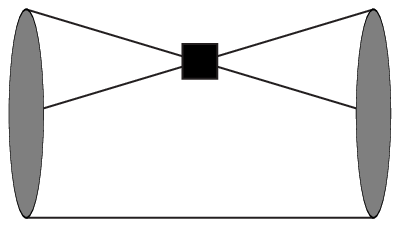,height=20cm}}
\vspace{-10cm}
\caption{}
\end{figure}

\begin{figure}
\vspace{-5cm}
\centering{\epsfig{figure=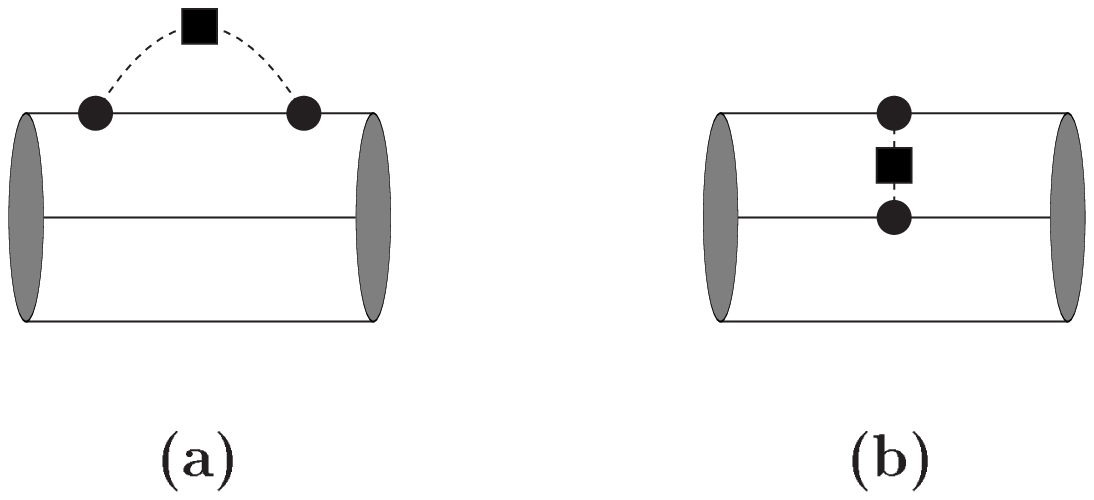,height=20cm}}
\vspace{-10cm}
\caption{}
\end{figure}

\begin{figure}
\vspace{-3cm}
\centering{\epsfig{figure=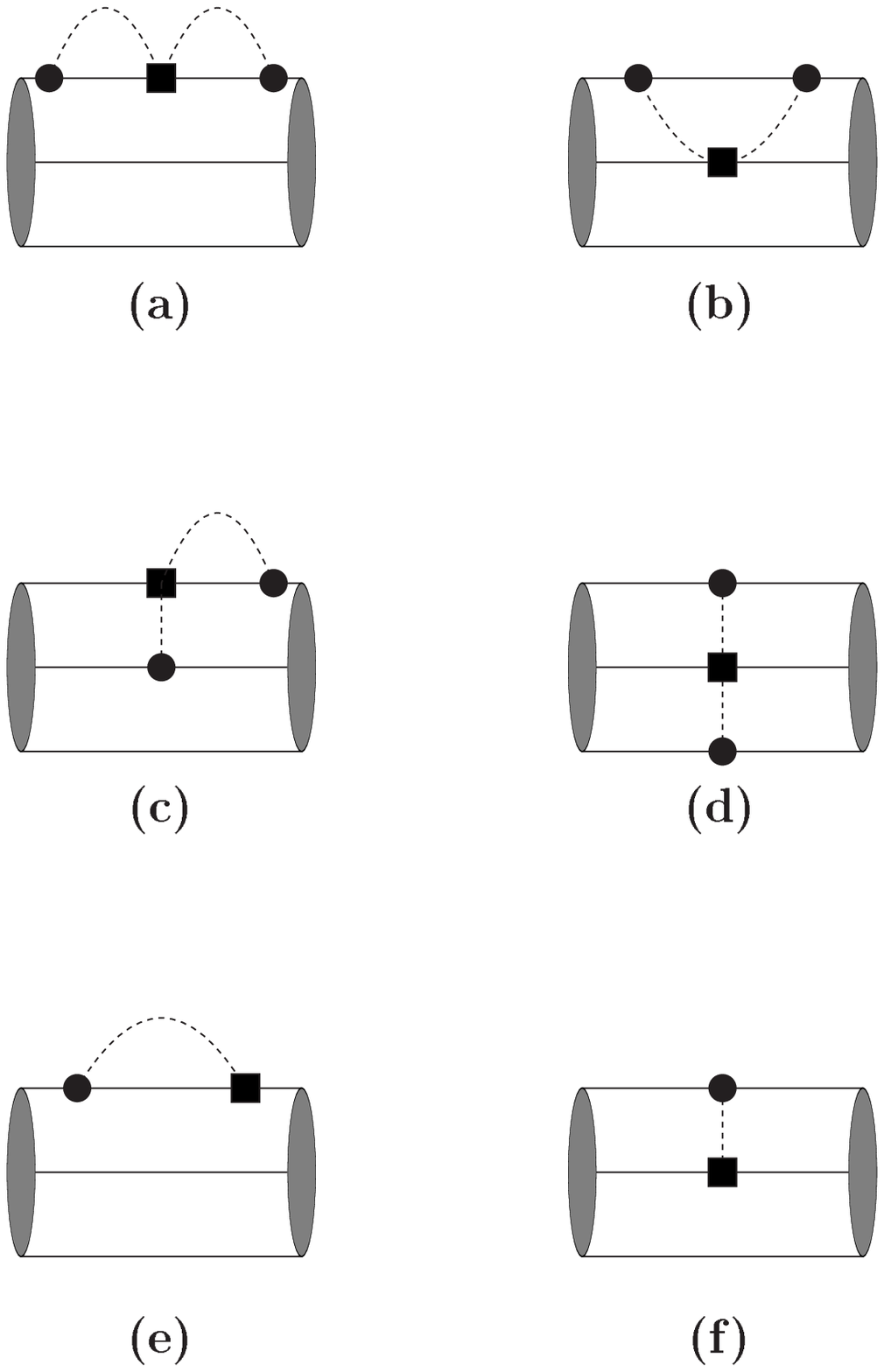,height=20cm}}
\vspace{-5cm}
\caption{}
\end{figure}

\end{document}